\documentclass[a4paper,11pt]{article}
\pdfoutput=1 

\usepackage{jheppub} 

\usepackage{xcolor}
\usepackage[font=footnotesize]{caption}
\usepackage{float}
\usepackage{braket}
\usepackage{extarrows}
\usepackage{bbm}
\usepackage{nicefrac}
 \usepackage{import}
 \usepackage{bbold}
\usepackage{mathrsfs}
\usepackage{pgf,tikz}
\usetikzlibrary{arrows}
\usepackage[fleqn]{mathtools}
\usepackage{graphicx}
\usepackage{subcaption}
\usepackage{dsfont}
\usepackage{enumerate}
\usepackage{enumitem}
\usepackage{nccmath}
\usepackage{booktabs}
\usepackage{siunitx}
\usepackage{tikz-cd}
\usepackage[parfill]{parskip}

\usepackage{graphicx, float, array, xspace, amscd, amsmath, amsthm, amssymb, latexsym, longtable,bbm}
\usepackage[bbgreekl]{mathbbol}
\usepackage{amsfonts}
\usepackage{url}
\usepackage{tikz}
\usepackage{lscape}

\usetikzlibrary{arrows.meta}
\usetikzlibrary{shapes.geometric}
\usetikzlibrary{positioning}

\newcommand{\X}{\Xi}

\DeclareFontFamily{OT1}{pzc}{}
\DeclareFontShape{OT1}{pzc}{m}{it}{<-> s * [1.200] pzcmi7t}{}
\DeclareMathAlphabet{\mathpzc}{OT1}{pzc}{m}{it}

\newcommand{\cC}{\mathcal{C}}

\newcommand{\cF}{\mathcal{F}}

\newcommand{\cK}{\mathcal{K}}

\newcommand{\cM}{\mathcal{M}}
\newcommand{\cN}{\mathcal{N}}

\newcommand{\cP}{\mathcal{P}}

\newcommand{\cT}{\mathcal{T}}

\newcommand{\cW}{\mathcal{W}}
\newcommand{\cX}{\mathcal{X}}

\newcommand{\IC}{\mathbb{C}}

\newcommand{\IQ}{\mathbb{Q}}
\newcommand{\IN}{\mathbb{N}}
\newcommand{\IR}{\mathbb{R}}

\newcommand{\IZ}{\mathbb{Z}}

\newcommand{\beq}{\begin{equation}}
\newcommand{\eeq}{\end{equation}}
\newcommand{\beqnn}{\begin{equation*}}
\newcommand{\eeqnn}{\end{equation*}}
\newcommand{\bea}{\begin{eqnarray}}
\newcommand{\eea}{\end{eqnarray}}
\newcommand{\bean}{\begin{eqnarray*}}
\newcommand{\eean}{\end{eqnarray*}}

\newcommand{\nn}{\nonumber}

\newcommand{\place}[3]{\vbox to0pt{\kern-\parskip\kern-7pt
                             \kern-#2truein\hbox{\kern#1truein #3}
                             \vss}\nointerlineskip}

\renewcommand{\Re}{\text{Re~}}

\newcommand{\tr}{\,\text{Tr}\,}

\usepackage{lscape}

\newenvironment{gleichung*}{\begin{equation*}\begin{aligned}}{\end{aligned}\end{equation*}}
\renewcommand{\sc}{\textsc}
\renewcommand{\bf}{\textbf}

\newcommand{\dd}{\mathrm{d}}
\newcommand{\cTmapsto}{\xmapsto{\cT}}

\newcommand{\cPdmapsto}{\xmapsto{\cP_d}}
\newcommand{\Tmapsto}{\xmapsto{T}}
\newcommand{\Pmapsto}{\xmapsto{P}}

\newcommand{\cCP}{\cC \cP}

\newcommand{\txi}{\tilde{\xi}}

\newcommand{\tF}{\tilde{F}}
\newcommand{\indHVI}{\Xi}
\newcommand{\indHVII}{\Omega}

\newcommand{\fatdot}{{\boldsymbol{\cdot}}}

\newcommand{\ot}{\boldsymbol{t}}
\newcommand{\bF}{\boldsymbol{F}}
\newcommand{\bX}{\boldsymbol{X}}
\newcommand{\bcN}{\boldsymbol{\cN}}

\newcommand{\dvol}{{\rm vol}}

\DeclareMathOperator{\reP}{Re}
\DeclareMathOperator{\re}{Re}
\DeclareMathOperator{\imP}{Im}

\title{\boldmath  Time reversal and $\boldsymbol{CP}$ invariance in Calabi-Yau compactifications}

\author[ab]{Kilian B\"onisch,}
\author[c]{Mohamed Elmi,}
\author[d]{Amir-Kian Kashani-Poor,}
\author[be]{Albrecht Klemm}

\affiliation[a]{Max-Planck-Institut f\"ur Mathematik, Bonn, D-53111, Germany}
\affiliation[b]{Bethe Center for Theoretical Physics, Universit\"at Bonn, D-53115, Germany}
\affiliation[c]{NHETC and Department of Physics and Astronomy, Rutgers University, NJ, 08855, USA}
\affiliation[d]{LPENS, ENS, Université PSL, CNRS, Sorbonne Universit\'{e}, Université Paris Cité, F-75005 Paris, France}
\affiliation[e]{Hausdorff Center for Mathematics, Universit\"at Bonn, D-53115, Germany}
\affiliation[~]{}

\emailAdd{kilian@mpim-bonn.mpg.de} 
\emailAdd{elmi@physics.rutgers.edu}
\emailAdd{amir-kian.kashani-poor@ens.fr}
\emailAdd{aklemm@th.physik.uni-bonn.de}

\abstract{We revisit the question of time reversal and $CP$ invariance in Calabi-Yau compactifications. We show that time reversal invariance is respected by quantum corrections to the prepotential. In particular, field independent $\theta$ angles whose presence is dictated by requiring integrality of relevant monodromy transformations can take precisely the quantized values compatible with time reversal invariance. Furthermore, monodromy symmetry enlarges the region on moduli space on which time reversal is not spontaneously broken. We define the action of the $CP$ transformation for multi-parameter models and argue that on the slice of moduli space where it is defined, $CP$ is trivially a symmetry of the theory. For supersymmetric vacua that lie in this slice, we derive a condition on the third cohomology of the compactification manifold which determines whether $CP$ preserving fluxes exist that stabilize the moduli to such points. In the case of one-parameter models, the condition is always satisfied.}

\begin{document}
\rightline{
BONN-TH-2022-10
}
\maketitle

\flushbottom


\newpage
\section{Introduction}
\label{sec:intro}
Calabi-Yau compactifications of type II string theory (in)famously exhibit a moduli space of vacua. A vast amount of work has been invested towards devising mechanisms of inducing a potential on this space, with the aim of obtaining phenomenologically more realistic string theory models. A complementary approach towards lifting the degeneracy of vacua is to search for points on moduli space distinguished geometrically and then ask for physical implications of these distinguishing features. Supersymmetric flux vacua of type IIB were not historically, but could have been discovered following this strategy. In this paper, we revisit the realization of the discrete symmetries $T$ (time reversal) and $CP$ (charge conjugation and parity) in type II Calabi-Yau compactifications from this vantage point.

The question of time reversal invariance was recently discussed in the context of rigid Calabi-Yau compactifications in \cite{Cecotti:2018ufg}. There, a single $\theta$ angle appears, the coefficient of the topological term involving the graviphoton field strength. This coefficient is a field-independent constant, and the authors of reference  \cite{Cecotti:2018ufg} ask, in the spirit of the swampland program \cite{Vafa:2005ui}, whether it takes a distinguished value preserving time reversal invariance in rigid Calabi-Yau compactifications. Our analysis, in the broader context of general Calabi-Yau compactifications, differs qualitatively as the multiple $\theta$ angles that appear are field dependent. We address three questions: first, we show that the perturbative and non-perturbative corrections to the tree-level prepotential around the large radius point, which can be determined precisely via mirror symmetry, do not break time reversal invariance. Surprisingly, it is not the infinitely many instanton corrections which pose the bigger challenge, but subtle quadratic terms in the prepotential which are needed to ensure integral monodromy. We show that the quantized value of the coefficients of these terms lead precisely to the two values of $\theta$ angles which are compatible with time reversal symmetry. Secondly, we argue that the monodromy action on the period vector associated to the Calabi-Yau compactification extends the set of vacuum expectation values of non-invariant fields compatible with time reversal invariance away from zero. Finally, inspired by \cite{Cecotti:2018ufg} and in the spirit of the opening paragraph of this introduction, we ask whether the $\theta$ angles take interesting values at other distinguished points in moduli space. We note that rank~2 attractor points have the distinguishing feature that, with an important caveat that we discuss, the gauge coupling matrix decouples the graviphoton from the remaining vector fields. Encouraged by this result, we considered the explicit value of the complex graviphoton coupling at such a point in an example: the value is mathematically distinguished by its relation to periods of modular forms, but its physical relevance remains elusive. 

Considerations of $CP$ symmetry in the context of string theory date back to the early days of the field \cite{Strominger:1985it,Dine:1986bg,Dine:1992ya}. $CP$ must of course be broken in any realistic string model in order to reproduce the weak sector of the Standard Model. One focal point of the body of work on $CP$ in string theory is how to mitigate this breaking in the strong sector, i.e.\ how to solve (or incorporate solutions to) the strong $CP$ problem. A typical approach is to assume that $CP$ is broken in the underlying theory, and to attempt to calculate (or in recent works, determine the statistics) of the instanton generated potential for the Peccei-Quinn axion (an incomplete selection of such works is \cite{Conlon:2006tq,Svrcek:2006yi,Cicoli:2012sz,Broeckel:2021dpz, Demirtas:2021gsq}). In this work, we ask, in the context of flux compactifications of type IIB string theory, to what extent ingredients in these models {\it preserve} $CP$. After generalizing to multi-parameter models an old proposal of Strominger and Witten \cite{Strominger:1985it} for defining $CP$ transformations in the context of Calabi-Yau compactifications, we argue that $CP$ is trivially realized for Calabi-Yau compactifications on the locus of moduli space on which it can be defined, as it is induced by a orientation preserving diffeomorphism of the 10 dimensional theory (this is similar to an argument presented in \cite{Dine:1992ya}). We next discuss the vector and hyperscalar VEVs compatible with $CP$ invariance. For the supersymmetric vacua which preserve $CP$, we ask whether $CP$ invariant fluxes can be chosen which stabilize the moduli at these points. Similar questions have also been pursued in \cite{Kobayashi:2020uaj,Ishiguro:2020nuf,Ishiguro:2021ccl}, though we arrive at somewhat different conclusions: we argue that the fully corrected prepotential preserves $CP$ invariance, and we derive a condition on the third cohomology of the Calabi-Yau manifold which determines whether a supersymmetric flux vacuum preserves $CP$ symmetry. In the case of one-parameter models, we show that the condition is always satisfied.

From a four dimensional quantum field theory point of view, studying $T$ and $CP$ invariance separately is redundant, as based either on arguments relying on analytical continuation to Euclidean spacetime (see e.g.\ \cite{Streater:1989vi}) or on a detailed analysis of the types of interactions which can occur in a Lorentz invariant Lagrangian theory (see e.g.\ \cite{Weinberg:1995mt}), such theories enjoy $CPT$ invariance. In the larger context of higher dimensional quantum gravity theories, the two transformations appear on different footing: while the $T$ transformation can always be formulated, the existence of a natural candidate for a higher dimensional $CP$ transformation depends on the details of the compactification manifold. It thus makes sense to discuss the two independently.

The paper is organized as follows. In section \ref{sec:discrete_symmetries}, we discuss the basic structure of operators associated to discrete Lorentz symmetries, and identify two classes of terms in the action based on their transformation behavior under such symmetries. Section \ref{sec:time_reversal} discusses the action of time reversal in Calabi-Yau compactifications. As time reversal is orientation reversing, type IIA is the natural setting for this discussion. We identify a choice of intrinsic phases that renders the 10d action invariant in subsection \ref{subsec:T_in_10d}, before turning to the compactified theory in four dimensions in subsection \ref{subsec:T_in_4d}. Given the 10d result, the tree level theory must satisfy time reversal symmetry, as we check explicitly in \ref{subsubsec:T_in_4d_pert}. We incorporate non-perturbative $\alpha'$ corrections into our discussion in \ref{subsubsec:T_in_4d_inst}, and show that these respect time reversal invariance. In addition to worldsheet instanton contributions, mirror symmetry requires a constant and quadratic contributions to the prepotential. The coefficients of the latter are quantized and map to field independent $\theta$ angles. We work out the normalization of the action and show that the values that these coefficients may take are precisely those at which time reversal invariance holds. In subsection \ref{subsec:spontaneous_breaking_of_T}, we argue that while time reversal invariance seemingly requires the vacuum expectation value of the scalars in vector multiplets to vanish, VEVs equal to integers or half integers also preserve time reversal invariance as a consequence of monodromy symmetry.  Finally, in subsection \ref{subsec:T_away_from_large_radius}, we discuss our implicit assumption that the compactification takes place in a vicinity of the large radius point, and touch upon issues that arise when moving away from this point. We turn to the discussion of $CP$ invariance in section \ref{sec:CP}. Unlike time reversal, it is natural to define $CP$ so that it acts on the internal manifold. We discuss this action in section \ref{subsec:CP_internal}. As the combined action of $CP$ on spacetime and the internal manifold is orientation preserving, the invariance of the 10d theory follows from the analysis of section \ref{subsec:trans_p_forms} without the need to introduce intrinsic phases. It is however possible to introduce intrinsic phases, and this will prove useful in discussing flux vacua. This is discussed in section \ref{subsec:CP_10d_action}. The question of spontaneous breaking of $CP$ invariance is treated in section \ref{subsec:spontaneously_breaking_CP}. Finally, in section \ref{subsec:CP_flux_vacua}, we analyze the invariance of supersymmetric type IIB flux vacua under $CP$ transformations. A series of appendices complement the text. In appendix \ref{app:4d_SUGRA}, we review two aspects of 4d $\cN=2$ supergravity theories: the gauge coupling matrix $\cN$ in light of special geometry, and the symplectic invariance and monodromy symmetry of such theories. Appendix \ref{app:specialkaehler} reviews in some detail the special K\"ahler geometry of the complex structure moduli space of Calabi-Yau manifolds. We review supersymmetric flux vacua in the context of type IIB flux compactifications in appendix \ref{appendix:SUSY_vacua}. Appendix \ref{appendix:Finding rank 2 attractors} finally discusses how to explicitly find rank 2 attractor points, which are equivalent to supersymmetric vacua in one-parameter models. We provide a list of such points in table~\ref{tab:attractor}.

\section{Discrete Lorentz symmetries} \label{sec:discrete_symmetries}

The Lorentz group in arbitrary dimensions $d$ exhibits four connected components. The component containing the identity is called the proper orthochronous Lorentz group. The other three components are obtained by acting by time reversal $\cT$,
\begin{equation}
    t \cTmapsto -t \,, \quad x^i \cTmapsto x^i \,,
\end{equation}
space inversion $\cP_d$\,,
\begin{equation}
    t \cPdmapsto t \,, \quad x^i \cPdmapsto -x^i \,,
\end{equation}
and their composition $\cT \cP_d$.

Quantum field theory already in four spacetime dimensions does not allow us to distinguish between the action of a discrete Lorentz symmetry such as $\cP$ or $\cT$ and a product of this action with a global internal symmetry (i.e.\ one not involving an action on spacetime), see e.g.\ the discussion in \cite{Weinberg:1995mt}. When descending from higher dimensions, we have even more freedom to define the action of these symmetries, as we can couple them with an involutive action of our choice on the internal dimensions.\footnote{Note that the action $x^i \mapsto -x^i$ cannot be defined generically in the internal dimensions; indeed, generically, global coordinates $x^i$ do not exist, and we may or may not be able to define an involution on the manifold. More on this later.} The composition of any such action with the reversal of time which is a symmetry of the theory merits the name $\cT$, just as the composition with the inversion of the three spatial dimensions which yields a symmetry merits the name $\cP$. We denote the corresponding operators on the Hilbert space of the theory as $T$ and $P$ respectively.

In canonical quantization, the construction of quantum fields relies on imposing proper transformation properties under proper orthochronous Lorentz transformations. The transformation under $P$ and $T$ can involve intrinsic phases whose relative values can be partially worked out by analyzing the structure of the fields. The textbook \cite{Weinberg:1995mt} is an excellent reference on such matters. This analysis leads e.g.\ to the statement that a fermion and an anti-fermion have opposite intrinsic parity, implying that mesons that are S-wave bound states, such as pions, are pseudo-scalars.

\subsection[Transformation of $p$-form fields under discrete Lorentz symmetries]{Transformation of \boldmath{$p$}-form fields under discrete Lorentz symmetries} \label{subsec:trans_p_forms}
The bosonic fields arising in 10d supergravities are the metric, the dilaton, and $p$-form fields.

We will assume that $\cP$ and $\cT$ are isometries of the metric. We will also assume that they leave the dilaton invariant, given that we do not expect their application to result in strong-weak dualities.  We hence turn to the study of the transformation properties of $p$-forms. In physics, we often have the coefficients of a differential $p$-form in mind when we speak of a $p$-form field. For example, we think of the four components of the photon field as the coefficients of a 1-form, transforming under parity as
\begin{equation}
    A_0(x) \Pmapsto A_0(\cP x) \,, \quad A_i(x) \Pmapsto - A_i(\cP x) \,.
\end{equation}
The transformation properties of $p$-form fields under $P$ and $T$ are however most succinctly described if we consider the $p$-form as a whole. Writing $A = A_\mu \dd x^\mu$, the above transformation becomes
\begin{equation}
    A(x)  \Pmapsto  (\cP^* A)(x) = A_\mu(\cP x) \cP^* \dd x^\mu  \,.
\end{equation}

More generally, any $p$-form field $C$ can carry an intrinsic sign, in addition to the pullback action,
\begin{equation} \label{eq:C_under_T_and_P}
    C(x) \Pmapsto \pm (\cP^*C)(x) \,, \quad C(x) \Tmapsto \pm (\cT^*C)(x) \,.
\end{equation}

Contributions of $p$-form fields to the action fall into two categories: kinetic terms which are metric dependent via the occurrence of the Hodge star and metric independent topological terms. We can subsume the discussion of the action of $\cP$ and $\cT$ on these terms under the study of their fate under the action of a general diffeomorphism $\phi: M \rightarrow M$ on spacetime $M$. In the case of topological terms, the transformation under pullback of the fields via the diffeomorphism $\phi$ is given by
\begin{eqnarray} \label{eq:action_top_term}
    \lefteqn{S_{\text{top}}[C_i] = \int_M \omega_{i_1} \wedge \ldots \wedge \omega_{i_n} }\\
    &&\mapsto S_{\text{top}}[\phi^*C_i] = \int_M \phi^*\omega_{i_1} \wedge \ldots \wedge \phi^*\omega_{i_n} = \int_M \phi^* (\omega_{i_1} \wedge \ldots \wedge \omega_{i_n} ) = \pm S_{\text{top}}[C_i] \,,\nn
\end{eqnarray}
where the forms $\omega_{i}$ denote either $p$-form potentials $C_i$ or the associated field strengths $F_i$. The final sign is positive for orientation preserving and negative for orientation reversing maps $\phi$. The second type of contribution takes the form
\begin{equation}
    S_{\text{kin}}[g,C] = \int_M \dd C \wedge * \dd C = \int_M \langle \dd C, \dd C \rangle_g \dvol_g \,.
\end{equation}
Assuming $\phi$ to be an isometry of the metric, such contributions transform as
\begin{eqnarray}
    \lefteqn{S_{\text{kin}}[\phi^*g, \phi^*C ]=S_{\text{kin}}[g, \phi^*C ] = }\\
    &&\int_M \langle \dd  \phi^*C, \dd  \phi^*C \rangle_g \dvol_g = \pm \int_M \phi^* \big(\langle \dd C, \dd C \rangle_g \dvol_g \big) = S_{\text{kin}}[g,C] \,.
\end{eqnarray}
In the penultimate step, we have invoked
\begin{equation}
    \phi^* \dvol_g = \pm \dvol_g \,,
\end{equation}
with the sign depending on whether $\phi$ is orientation preserving (plus sign) or reversing (negative sign).

We conclude that kinetic energy type contributions are invariant under any isometry (orientation preserving or not), whereas topological terms are invariant under any orientation preserving diffeomorphism.

In the simple case of electromagnetism, the kinetic term $F\wedge*F$ is thus invariant under any isometry, while the topological term $F \wedge F$ breaks the symmetry under orientation reversing transformations. Note that both terms are insensitive to the choice of the intrinsic sign displayed in \eqref{eq:C_under_T_and_P}.
The analysis becomes sensitive to this sign when couplings between different $C$-form fields exist, or in the presence of sources. A 1-form field coupled via a covariant derivative 
\begin{equation}
    D = \dd + i A
\end{equation}
will preserve $P$ if it transforms without sign (as does $\dd$), and it will preserve $T$ if it transforms with sign (i.e.\ with opposite parity compared to $\dd$),
\begin{equation}
    A(x) \Pmapsto (\cP^*A)(x)  \,, \quad A(x) \Tmapsto - (\cT^*A)(x) \,.
\end{equation}
(recall that only if $T$ is realized as an anti-linear operator can it relate two theories which both exhibit a bounded spectrum, as a linear $T$ would map $H \mapsto -H$). On the other hand, both signs are compatible with matter charged under shift symmetries, as occurs e.g.\ in flux compactifications.

Note that self-duality conditions such as
\begin{equation}
    F_5 = * F_5
\end{equation}
of type IIB supergravity are not compatible with orientation reversing isometries, as by
\begin{eqnarray}
    \lefteqn{\phi^*(\eta \wedge * \omega) = \phi^*\eta \wedge \phi^* (*\omega) =} \\
    &&\phi^* \big( \langle \eta, \omega \rangle_g \dvol_g \big) = \langle \phi^* \eta, \phi^* \omega \rangle_g \phi^* \dvol_g = \pm \phi^* \eta\wedge * \phi^* \omega \,,
\end{eqnarray}
an orientation reversing isometry $\phi$ anti-commutes with the Hodge star
\begin{equation} \label{eq:PhionHodgeStar}
    \phi^* (* \omega) = - * \phi^* \omega \,.
\end{equation}

\section{Time reversal} \label{sec:time_reversal}
In this section, we will explore the time reversal symmetry of type II string theory compactifications on Calabi-Yau manifolds, which lead to 4d theories with $\cN=2$ supersymmetry. The action of time reversal in 4d spacetime is orientation reversing. Unlike the case of parity to which we shall turn below, it does not appear natural to compose this action with an action on the internal dimensions in defining $T$. By the argument at the end of section \ref{subsec:trans_p_forms}, it is therefore difficult to take type IIB supergravity as a starting point for our considerations, and we anchor our discussion in type IIA theory instead. Note that by mirror symmetry, both 10d vantage points should ultimately give rise to the same conclusions in 4d.

\subsection{The action of time reversal in 10d supergravity} \label{subsec:T_in_10d}
The bosonic action of type IIA supergravity is given by
\begin{eqnarray}
    S^{\text{IIA}} &=& \frac{1}{2\kappa^2} \int \Big[ e^{-2\phi} \left(  R *1 + 4 \dd \phi \wedge * \dd \phi - \frac{1}{2} H_3 \wedge * H_3 \right) \label{eq:10d_action}\\
    && - \frac{1}{2}  \left( F_2 \wedge * F_2 + F_4 \wedge * F_4 \right)  - \frac{1}{2}  \left(B_2 \wedge \dd C_3 \wedge \dd C_3  \right) \Big] \,, \nonumber
\end{eqnarray}
where 
\begin{equation}
    F_2 = \dd C_1 \,, \quad F_4 = \dd C_3 - B_2 \wedge \dd C_1 \,, \quad H_3 = \dd B_2 \,.
\end{equation}
By the discussion in section \ref{subsec:trans_p_forms}, we need to introduce intrinsic phases under time reversal to render the topological terms in this action time reversal invariant. As the term
\begin{equation}
    B_2 \wedge \dd C_3 \wedge \dd C_3
\end{equation}
is quadratic in $\dd C_3$, it fixes the required transformation property
\begin{equation} \label{eq:TonB}
    B_2(x) \Tmapsto - \cT^*(B_2)(x) 
\end{equation}
of $B_2$ uniquely. But then, for $F_4$ to transform simply under time reversal, we need to require that $C_1$ and $C_3$ transform with opposite relative sign,
\begin{equation}
    C_1(x) \Tmapsto \pm (\cT^*C_1)(x) \,, \quad C_3(x) \Tmapsto \mp (\cT^* C_3)(x) \,. \label{eq:TonAandC}
\end{equation}

The $p$-form fields occurring in type II string theory are sourced by D-branes. The coupling occurs via a term 
\begin{equation} \label{eq:WZW_coupling}
   \mu \int_V \tr \left[ \exp[2\pi \alpha' \cF_2 + B_2] \sum_q C_q \right] 
\end{equation}
in the D-brane worldvolume action. Here, $V$ denotes the worldvolume of the brane, $\mu$ its tension and $\cF_2$ the field strength (which can be non-abelian, thence the trace) of the gauge field on the brane. From the form of the coupling \eqref{eq:WZW_coupling}, we can read off that in order to preserve $T$,
\begin{itemize}
    \item the field strength on the brane must transform as \eqref{eq:TonB},
    \item $C_{i}$ and $C_{i+2}$ must transform with opposite sign. This condition is consistent with \eqref{eq:TonAandC}. It also implies that $\dd C$ and $*\dd C$ transform with equal sign (as $C_1$ and $C_7$ are electric-magnetic duals, as are $C_3$ and $C_5$).
\end{itemize}
 

\subsection{The action of time reversal on the 4d theory}
\label{subsec:T_in_4d}
Having shown the invariance of the 10d supergravity action under time reversal, the invariance of the 4d theory obtained from it upon compactification is automatic. By invoking mirror symmetry, $\alpha'$ corrections to the theory can be computed and elegantly packaged at the level of the 4d theory. We will set the stage in the next subsection by verifying the time reversal invariance of the tree level 4d action, before turning to the $\alpha'$ corrected action in section \ref{subsubsec:T_in_4d_inst}.

\subsubsection{The theory at tree level} \label{subsubsec:T_in_4d_pert}
The 4d supergravity action obtained from type IIA upon Calabi-Yau compactification will inherit time reversal symmetry. We can see this explicitly. The bosonic action is equal to
\begin{equation} \label{eq:4d_action}
    S^{4\text{d}} = \int \Big[ \frac{1}{2} R*1 - g_{i\bar{\jmath}} \dd t^i \wedge * \dd\bar{t}^{\bar{\jmath}} - h_{uv}\dd q^u \wedge * \dd q^v + \frac{1}{2} \imP \cN_{IJ} F^I \wedge * F^J + \frac{1}{2} \reP \cN_{IJ} F^I \wedge F^J \Big]
\end{equation}
with the metric on the hypermultiplet moduli space given by
\begin{eqnarray}
    h_{uv} \dd q^u \wedge * \dd q^v &=& \dd \phi \wedge * \dd \phi + g_{a\bar{b}} \dd z^a \wedge * \dd \bar{z}^{\bar{b}} + \\
    && +\frac{e^{4\phi}}{4} \left(\dd a + \frac{1}{2} (\txi_A \dd \xi^A - \xi^A \dd \txi_A )\right) \wedge * \left(\dd a + \frac{1}{2}(\txi_A \dd \xi^A - \xi^A \dd \txi_A) \right) + \nn \\
    && -\frac{e^{2\phi}}{2} (\imP \cM^{-1})^{AB} \left( \dd\tilde{\xi}_A + \cM_{AC}\dd \xi^C \right) \wedge * \left( \dd\tilde{\xi}_A + \overline{\cM}_{AC}\dd\xi^C \right) \nn \,.
\end{eqnarray}
The index $i$ (as well as its alphabetic neighbors\footnote{This qualifier will also apply to all ensuing index attributions $I,a,A, \ldots$.}) enumerates vector multiplets containing each one complex scalar field $t^i$ and a vector field whose field strength is denoted $F^i$. The index $I$ runs over the range of $i$ with $0$ adjoined. $F^0$ is the field strength of the graviphoton, which resides in the $\cN=2$ gravity multiplet, together with the metric. The special geometry relations governing the vector multiplet sector are summarized in appendix \ref{app:4d_SUGRA}. The hypermultiplets are indexed by $A$, which runs over the range of $a$ with $0$ adjoined. The dilaton $\phi$, the axion $a$ and the real pair of scalars $(\xi^0 ,\tilde{\xi}_0)$ reside in the so-called universal hypermultiplet, while all other hypermultiplets combine a complex scalar field $z^a$ with a pair of real scalars $(\xi^a ,\tilde{\xi}_a)$.
The matrix ${\cM}$ is the mirror dual to the gauge coupling matrix $\cN$: in type IIA compactifications on a Calabi-Yau manifold $X$, its expression is given by \eqref{eq:NIJ}, with the prepotential occurring in this definition determined by the special geometry of the complex structure moduli space of $X$. Likewise, the K\"ahler metric $g_{a\bar{b}}$ on the special K\"ahler base of the hypermultiplet moduli space is given by \eqref{eq:sigma_model_metric}, based on the same prepotential.

We shall first consider the transformation behavior of the hyperscalars under time reversal. We can let $z^a$ and the dilaton transform trivially. The matrix $\cM$ as a function of $z^a$ is therefore also invariant. The transformation behavior of the axion $a$ is determined by that of the 10d $B$-field: it is related to the space-time components $h_3$ of the field strength of $B$ via
\begin{equation} \label{eq:dual_h3}
    \dd a = * h_3 + \ldots \,.
\end{equation}
By \eqref{eq:PhionHodgeStar} and \eqref{eq:TonB},
\begin{equation}
    *h_3(x) \Tmapsto - * (\cT^* h_3)(x) = \cT^*(*h_3)(x)  \,.
\end{equation}
Hence, the intrinsic phase of $a$ under time reversal is +1. Finally, the hyperscalars $\xi^A$ and $\tilde{\xi}_A$ in a type IIA compactification on $X$ arise as the expansion coefficients of $C_3$ in a symplectic basis of $H^3(X,\IZ)$ and therefore both transform with the same sign under time reversal. 

We conclude that the hypermultiplet sector conserves time reversal invariance at tree level, no matter what sign we choose in \eqref{eq:TonAandC}. 

The hypermultiplet sector generically receives $g_s$ corrections, yet is protected in type IIA compactifications against $\alpha'$ corrections. As one choice of sign in \eqref{eq:TonAandC} leaves the hypermultiplet sector untouched, we can rule out time reversal breaking contributions in the fully quantum corrected action as long as hyper- and vector multiplet contributions do not mix, i.e.\ up to two derivative level. We are tempted to conjecture that the quantum corrected action will retain the symmetry under $(\xi^A, \txi_A) \mapsto (-\xi^A, -\txi_A)$, to render our argument independent of the choice of sign in \eqref{eq:TonAandC}. 

Turning now to the more interesting vector multiplet sector, recall that the 10d origin of the graviphoton $A^0$ is the gauge potential $C_1$, and that the real part of the complex scalar fields $t^i$ residing in vector multiplets descend from internal modes of the 10d $B_2$ field, while the imaginary parts encode K\"ahler moduli of the internal metric,
\begin{equation}
    t^i = b^i + i v^i \,.
\end{equation}
The behavior of $b^i$ under time reversal follows from \eqref{eq:TonB}:
\begin{equation} \label{eq:time_reversal}
    b^i(x) \Tmapsto - b^i(\cT x) \,, \quad \mathrm{i.e.}\quad t^i(x) \Tmapsto -\overline{t^i}(\cT x) \,,
\end{equation}
as $T$ acts as an isometry on the metric. We lift the action of time reversal to projective coordinates on the vector multiplet moduli space via
\begin{equation} \label{eq:lift_of_T_to_X}
    X^0(x) \Tmapsto \pm \overline{X^0}(\cT x) \,, \quad X^i(x) \Tmapsto \mp \overline{X^i}(\cT x) \,.
\end{equation}

Dimensional reduction of the 10d action \eqref{eq:10d_action} leads to the 4d action \eqref{eq:4d_action} with the $\sigma$-model metric $g_{i \bar{\jmath}}$ and the gauge coupling matrix $\cN_{IJ}$ obtained from the cubic prepotential
\begin{equation} \label{eq:tree_level_prepotential}
    F^{\text{tree}} = - \frac{1}{3!} \frac{\kappa_{ijk} X^i X^j X^k}{X^0} \,.
\end{equation}
Here, $\kappa_{ijk}$ denote the triple intersection numbers, see \eqref{eq:triple_intersection}. The gauge coupling matrix which follows from this prepotential via equation \eqref{eq:NIJ} has components $\reP \cN_{00}$, $\reP \cN_{ij}$, $\imP \cN_{i0}$ which are odd in the fields $b^i$, and complementary components that are even. As $\reP \cN$ is the coefficient matrix of the topological term, and $\imP \cN$ the coefficient matrix of the gauge kinetic term, this is the 4d manifestation of the 10d argument leading to \eqref{eq:TonAandC}: the graviphoton must transform with opposite sign relative to all other gauge fields (which belong to vector multiplets) in order for time reversal to be a symmetry of the action.

\subsubsection[The $\alpha'$ corrected theory]{The \boldmath{$\alpha'$} corrected theory} \label{subsubsec:T_in_4d_inst}
The above discussion was for the tree level action obtained from dimensional reduction of the type IIA action. The vector multiplet sector is protected against $g_s$ corrections, but does receive $\alpha'$ corrections in type IIA compactifications. These are completely captured via mirror symmetry. We discuss these corrections in this subsection.

Note first that, as the $\kappa_{ijk}$ occurring in \eqref{eq:tree_level_prepotential} are real, the transformation of the tree level prepotential $F^{\text{tree}}$ under time reversal is given by
\begin{equation} \label{eq:T_transformation_of_prepotential}
    t^i(x) \Tmapsto - \overline{t^i}(\cT x) \quad \Rightarrow  \quad F(x) \Tmapsto - \overline{F}(\cT x) \,,
\end{equation}
where we have written $F$ for $F^{\text{tree}}$. In fact, independently of the precise form of the prepotential $F$, the behavior \eqref{eq:T_transformation_of_prepotential} alone guarantees invariance of the action under time reversal, as it implies that the components $\reP \cN_{00}$, $\reP \cN_{ij}$, $\imP \cN_{i0}$ change sign under time reversal, while the complementary components remain invariant. We can see this directly from the presentation \eqref{eq:NIJ} of the gauge coupling matrix in terms of the prepotential: writing
\begin{equation}
    F_I = \frac{\partial F}{\partial X^I} \,, \quad F_{IJ} = \frac{\partial^2 F}{\partial X^I \partial X^J} \,,
\end{equation}
note that \eqref{eq:T_transformation_of_prepotential} together with \eqref{eq:lift_of_T_to_X} imply
\begin{equation}
    F_0(x) \Tmapsto \mp \overline{F_0}(\cT x) \,, \quad F_i(x) \Tmapsto \pm \overline{F_i}(\cT x)
\end{equation}
and
\begin{equation}
    F_{00}(x) \Tmapsto -\overline{F_{00}}(\cT x) \,, \quad F_{i0}(x) \Tmapsto \overline{F_{i0}}(\cT x) \,, \quad F_{ij}(x) \Tmapsto - \overline{F_{ij}}(\cT x) \,,
\end{equation}
from which the claim easily follows. Alternatively, we can begin by considering one of the defining relations for $\cN_{IJ}$ given in \eqref{eq:NIJ_bis1},
\begin{equation}
    \cN_{IJ} X^J = F_I \,.
\end{equation}
Then
\begin{equation}
\begin{tikzcd}
&\cN_{0J} X^J/X^0 (x)\arrow[mapsto]{d}[swap]{T} \arrow[r, equal] 
    & F_0/X^0 (x)\arrow[mapsto]{d}[swap]{T}\\
    & (\tilde{\cN}_{00} - \tilde{\cN}_{0j} \bar{t}^j)(\cT x) \arrow[r,equal] 
    & -\overline{F_0}/\overline{X^0} (\cT x)\arrow[r, equal]
    &- (\overline{\cN}_{00} + \overline{\cN}_{0j}\bar{t}^j)(\cT x)
\end{tikzcd} 
\end{equation}
and 
\begin{equation}
\begin{tikzcd}
&\cN_{iJ} X^J/X^0 \arrow[mapsto]{d}[swap]{T} \arrow[r, equal] 
    & F_i/X^0 \arrow[mapsto]{d}[swap]{T}\\
    & (\tilde{\cN}_{i0} - \tilde{\cN}_{ij} \bar{t}^j)(\cT x) \arrow[r,equal] 
    & \overline{F_i}/\overline{X^0} (\cT x)\arrow[r, equal]
    & (\overline{\cN}_{i0} + \overline{\cN}_{ij}\bar{t}^j)(\cT x)
\end{tikzcd}
\end{equation}
where $\tilde{\cN}_{IJ}(\cT x)$ indicates the image of $\cN_{IJ}(x)$ under time reversal. Comparing the constant terms and coefficients of $\bar{t}^i$ yields the result.

This generalization away from $F^{\text{tree}}$ is important, as in 4d, non-perturbative quantum corrections can be elegantly packaged at the level of the action in terms of corrections to the tree-level prepotential. Based on the foregoing discussion, we conclude that these corrections will not break time reversal invariance if the corrected prepotential still transforms according to \eqref{eq:T_transformation_of_prepotential}. In the vicinity of the large radius point, the exact prepotential has the form
\begin{equation} \label{eq:exact_prepotential}
    \cF(\boldsymbol{t}) =-\frac{\kappa_{ijk}}{6} t^i t^j t^k -\frac{\sigma_{ij}}{2}t^i t^j+  \gamma_j t^j + \frac{\zeta(3) \chi }{2 (2 \pi i)^3}- \frac{1}{(2\pi i)^3} \sum_{\boldsymbol{n}} a_{\boldsymbol{n}} e^{2\pi i \boldsymbol{n\cdot t}} \,, \quad a_{\boldsymbol{n}} \in \IQ \,,
\end{equation}
where $(X^0)^2\cF(\boldsymbol{t}) = F(\boldsymbol{X})$. The coefficients appearing in this expansion are explained in the appendix following equation \eqref{eq:large_radius_F_multi_parameter}. The non-perturbative contribution satisfies \eqref{eq:T_transformation_of_prepotential} by reality of the coefficients $a_{\boldsymbol{n}}$. The perturbative contribution, polynomial in $t^i$, satisfies \eqref{eq:T_transformation_of_prepotential} if the coefficients of all odd order terms in the variables $t^i$ are real, and the coefficients of all even order terms imaginary. This is the case for all models for which the matrix $(\sigma)_{ij} = 0$. Before concluding that models with some entries $\sigma_{ij} \in \frac{1}{2}\IZ - \{0\}$ (the only other values that $\sigma_{ij}$ can take, see appendix~\ref{app:specialkaehler}) break time reversal invariance, we should note that by reality of the coefficients $\sigma_{ij}$, the order 2 terms in the prepotential only contribute linearly to the real part of the gauge coupling matrix \eqref{eq:NIJ}, giving rise to a term
\begin{equation} \label{eq:bare_theta}
    -\frac{1}{2} \sigma_{ij}F^i \wedge F^j
\end{equation}
in the action, up to a normalization constant we have not been keeping track of up to this point. As the values of $\sigma_{ij}$ can shift by integral amounts under monodromy, this coupling is only well-defined if we can identify $2\pi \sigma_{ij}$ with the $\theta_{ij}$ angle of periodicity $2\pi$. If this is true, $\sigma_{ij} \in \frac{1}{2} \IZ$ is exactly the constraint which ensures that $\int F^i \wedge F^j \mapsto - \int F^i \wedge F^j$ is a symmetry of the action.

To check this claim, we need to work out the correct normalization of the term \eqref{eq:bare_theta} by reinstating all dimensionful constants and keeping track of the integrality properties of the gauge fields. To keep constants such as $\kappa_{ijk}$ dimensionless, it will be convenient, deviating from standard practice in 4d, to not assign mass dimension to coordinates $x^i$ and differentials $\dd x^i$. The correct mass dimension $\ell^8$ of the topological term in the 10d action \eqref{eq:10d_action}, which requires
\begin{equation}
    [B_2] = [H_3] = \ell^2 \,, \quad[C_p] = [F_{p+1}] = \ell^p \,,
\end{equation}
then follows from assigning the appropriate mass dimension to the fields $(B_2)_{\mu \nu}$ and  $(C_p)_{\mu_1 \cdots \mu_p}$ rather than to the differentials $\dd x^{\mu_i}$. To ensure the correct mass dimension of the kinetic terms, we must assign
\begin{equation}
[g_{\mu \nu} ] = \ell^2 
\end{equation}
such that 
\begin{equation}
    [*1] = [\sqrt{-g}] = \ell^{10}
\end{equation}
and
\begin{equation}
[F_p \wedge * F_p] = [\frac{\sqrt{-g}}{p!} F_{\mu_1 \ldots \mu_p} F_{\nu_1 \ldots \nu_p} g^{\mu_1 \nu_1} \cdots g^{\mu_p \nu_p}] = \ell^{10} \ell^{2(p-1)} \ell^{-2p} = \ell^8 \,.
\end{equation}
After these preliminaries, we turn to the integrality properties of the form fields. The relation between the field strengths $F_p$ and integral cohomology follows from the Dirac quantization condition associated to the D-brane source term for these fields (see e.g.\ \cite{Blumenhagen:2013fgp}, also for the following statements relating couplings to the string length $l_s$):
\begin{equation} \label{eq:integrality_fluxes}
    \mu_{p-1} \int_{\Sigma_{p+1}} F_{p+1} \in 2 \pi \IZ \,.
\end{equation}
The BPS condition implies that the charge $\mu_{p-1}$ of a D$(p-1)$ brane equals its tension $T_{p-1}$. Assuming that the tension of the fundamental string and a D$1$ brane coincide, a worldsheet calculation yields
\begin{equation}
    T_p = \frac{2\pi}{l_s^{p+1}} \,,
\end{equation}
where $l_s^2 = 4 \pi^2 \alpha'$. It follows that (with apologies for the multiple uses of the bracket $[\cdot]$)
\begin{equation} \label{eq:flux_quantization}
    \left[\frac{F_{p+1}}{l_s^p} \right] \in H^{p+1}(X,\IZ) \,.
\end{equation}
A similar worldsheet calculation also yields the 10d gravitational coupling $\kappa^2$ in terms of the string length,
\begin{equation}
    \kappa^2 = \frac{1}{4\pi} l_s^8 \,.
\end{equation}
Next, we reinstate the $\alpha'$ dependence in the relation between the field $B_2$ appearing in \eqref{eq:10d_action} and the variable $b^i$ on which the prepotential \eqref{eq:exact_prepotential} depends via $t^i = b^i + i v^i$. The exponentials 
\begin{equation}
    e^{2 \pi i \boldsymbol{n \cdot t}}
\end{equation}
arise from worldsheet instantons. The $b^i$ dependence stems from the worldsheet integrals
\begin{equation}
    \exp\left({\frac{i}{4\pi \alpha'} \int_\Sigma B}\right) \,.
\end{equation}
Introducing the notation $b^i_s$ for the modes of $B$ (in the string worldsheet normalization), we conclude
\begin{equation}
    b^i = \frac{1}{8 \pi^2 \alpha'} b^i_s = \frac{b^i_s}{2l_s^2}  \,.
\end{equation}
We are now ready to perform the reduction of the topological term
\begin{equation}
    -\frac{1}{4\kappa^2} \int B_2 \wedge \dd C_3 \wedge \dd C_3
\end{equation}
in the action \eqref{eq:10d_action}, which leads to the perturbative contribution proportional to $\Re \cN_{ij}$ in the 4d action: choosing a basis of representatives $\{\omega_i\}$ of the cohomology $H^2(X,\IZ)$ normalized as
\begin{equation}
    \int_X \omega_i \wedge \omega_j \wedge \omega_k = \kappa_{ijk} \,,
\end{equation}
we obtain
\begin{eqnarray}
    -\frac{1}{4\kappa^2} \int_{M_4 \times X} B_2 \wedge \dd C_3 \wedge \dd C_3 &=& - \frac{\pi}{l_s^8} \int \kappa_{ijk} b^i_s \,\dd A_s^j \wedge \dd A_s^k \\
    &=& - \frac{2\pi}{4\pi^2} \int \kappa_{ijk} \,\frac{b^i_s}{2 l_s^2}\, \dd\,\frac{2 \pi A_s^j}{l_s^3} \wedge \dd\frac{2\pi A_s^k}{l_s^3} \\
    &=& - \frac{2\pi}{4\pi^2} \int \kappa_{ijk} \,b^i \,\dd A^j \wedge \dd A^k \,.
\end{eqnarray}
To accompany $b^i_s$, we have here introduced the modes $C_3 = A^i_s \omega_i + \ldots$, which by \eqref{eq:integrality_fluxes} are related to gauge fields with the conventional 4d normalization $\int F \in 2 \pi \IZ$ via
\begin{equation}
    A^i = \frac{2\pi}{l_s^3} A_s^j \,.
\end{equation}
Comparing to the corresponding term in the 4d action \eqref{eq:4d_action},
\begin{equation}
    \frac{1}{2} \int \reP \cN_{ij} F^i \wedge F^j \,,
\end{equation}
we conclude
\begin{equation}
    \frac{1}{2} \reP \cN_{ij}^{\text{tree}} = - \frac{2 \pi}{4 \pi^2} \, \kappa_{ijk}b^k \,. 
\end{equation}
With the correct normalization thus fixed, $\sigma_{ij}$ hence contributes
\begin{equation}
    \frac{1}{2}\reP \cN_{ij}^{\sigma} = -\frac{2\pi}{4\pi^2} \sigma_{ij}
\label{eq:sigmaoneparameter}
\end{equation}
to the gauge kinetic term, as we wished to show.

Note that in non-abelian gauge theories whose gauge group exhibits a non-trivial center, the physics at $\theta=0$ and $\theta = \pi$ is markedly different \cite{Gaiotto:2017yup}. Such theories are accessible via Calabi-Yau compactifications of type II string theory via the process of geometric engineering \cite{Katz:1996fh}, and thus fit into the framework just described. As an example of this setup, we consider the engineering of gauge theories with the same gauge group SU(2) and vanishing matter content but different $\theta$ angle. The engineering geometries are Calabi-Yau threefolds which can be presented both as K3 fibrations and as elliptic fibrations over the Hirzebruch surfaces $F_n$, $n=0,1$.\footnote{Over the base $F_1$, the K3 and the elliptic fibration are associated to different K\"ahler cones in the extended K\"ahler cone of the Calabi-Yau manifold.} These compact Calabi-Yau manifolds exhibit 
three K\"ahler moduli, traditionally labelled by $s$, $t$, and $u$: the first corresponds to the base $[B]$ of the rationally fibered  Hirzebruch surface, the second to its rational fiber $[F]$, and the third corresponds to the class $[F]+[E]$, with $[E]$ the class of the 
elliptic fiber. In terms of these parameters, the prepotential reads (in the K\"ahler cone of the K3 fibration)
\begin{eqnarray}
{\cal F}&=&-\frac{1}{6}(8\, u^3+3(2+n)\, u^2 s+ 6\, u^2 t+3n \,u t^2 +6 \,stu) \nn\\
&&- \frac{ n }{2} u t +\frac{1}{24}(92\, u+12(2+n)\, t +24\,s) + \frac{480 \zeta(3)}{2(2\pi i) ^3}+{\cal F}_{\text{inst}}(Q)\ .
\label{eq:Fstu}
\end{eqnarray}   
The form of the coefficient of the quadratic term $ut$ follows upon imposing integrality of the transformation matrix implementing the monodromies of the period vector under $u\mapsto u+1$ and $s\mapsto s+1$, as we argue in appendix \ref{app:computing_omega} after equation \eqref{eq:TshiftsatMUM}. In all three cases,  
we can decouple gravity by taking the volume of the elliptic curve 
$[E]$ to infinity, while performing a double scaling limit on the remaining two K\"ahler classes in which we also take 
the volume of the base $[B]$ to infinity, while the volume of the fiber
$[F]$, which governs the masses of the $W^\pm$ bosons of the SU(2) gauge theory, becomes hierarchically 
small~\cite{Katz:1996fh}. Note that this is the weak coupling limit of the dual heterotic 
string, as the volume $s\sim 1/g^2_{\text{het}}$ of the base $[B]$ of the Hirzebruch surfaces can be identified with the base of the K3 fibration. Following the 
analysis of~\cite{Katz:1996fh}  we  see from the classical terms in \eqref{eq:Fstu} that in the decoupling 
limit the real part of the gauge kinetic function evaluates  to $\frac{1}{2} {\rm Re } {\cal N}^\sigma_{tt}=- 
\frac{1}{4 \pi} n $, where $t$ corresponds to the scalar vacuum expectation value of the U(1) vector multiplet inside the SU(2). This indicates that we can engineer SU(2) Seiberg-Witten theory with $\theta=0$ for $n=0$ and $\theta = \pi$ for $n=1$.

\subsection{Spontaneously breaking time reversal invariance} \label{subsec:spontaneous_breaking_of_T}
In the previous subsection, we concluded that time reversal acts on the field $\ot(x)$ as 
\begin{equation}
    \ot(x) \Tmapsto - \overline{\ot}(\cT x) \,.
\end{equation}
The vacuum expectation value $\ot_0$ of the field $\ot$ is invariant under this transformation only if (assuming $\ot_0$ constant)
\begin{equation}
    \re \ot_0 = 0 \,.
\end{equation}
Before concluding that all other VEVs break time reversal symmetry, we recall that $\cN=2$ supergravity permits a symplectic action in the vector multiplet sector, as we review in appendix \ref{app:sympl_vs_monodromy}. A subgroup of the symplectic group acts as a symmetry on the theory. In type IIA Calabi-Yau compactifications, this symmetry group can be identified with the monodromy group acting on the middle dimensional homology of the mirror Calabi-Yau manifold. As long as we do not consider non-trivial gauge field backgrounds, we can identify all VEVs of $\ot$ related by this symmetry as describing the same theory.\footnote{In the presence of non-trivial gauge field backgrounds, the identification would also require transforming this background.} The monodromy around the MUM point of a Calabi-Yau manifold can be written down in terms of its topological invariants, see \eqref{eq:TshiftsatMUM}; it induces the following shift symmetries on the variable $\ot$:
\begin{equation}
    \ot \rightarrow \ot + \sum_i n_i \boldsymbol{e}_i \,, \quad n_i \in \IZ \,.
\end{equation}
Given $\ot_0$ such that $\re t_0^i \in \{0,\pm \frac{1}{2}\}$, the choice 
\begin{equation}
    n_i = 
    \begin{cases}
        -1 \quad &\text{for } \re t_0^i = \frac{1}{2} \\
        \hphantom{-}0 \quad &\text{for } \re t_0^i = 0 \\
        \hphantom{-}1 \quad &\text{for } \re t_0^i = -\frac{1}{2} \\
    \end{cases}
\end{equation}
acts just as negative complex conjugation. We conclude that all VEVs $\re t_0^i \in \{0,\pm \frac{1}{2}\}$ are compatible with the conservation of time reversal invariance.

\subsection{Time reversal symmetry away from the large radius point} \label{subsec:T_away_from_large_radius}
Up to now, we have implicitly considered time reversal symmetry in a vicinity of the large radius point. It is in this region that the 4d action \eqref{eq:4d_action}, with $h^{1,1}(X)$ vector multiplets, $h^{2,1}(X)+1$ hypermultiplets, and the prepotential given by \eqref{eq:exact_prepotential}, is a valid approximation of the theory. In particular, we have full knowledge of the massless spectrum of the theory here. Moving away from the large radius point, two phenomena may occur which invalidate the action \eqref{eq:4d_action}: additional states may become light, and a non-perturbative symplectic transformation may be required which changes which entries in the period vector \eqref{eq:period_vector} can be chosen to define coordinates on the scalar manifold of the vector multiplet sector (see the discussion in appendix \ref{app:sympl_vs_monodromy}). As the computation of the period vector takes place on the mirror manifold, it is convenient to continue the discussion from the vantage point of the mirror, and we will do so for the rest of this section. 

Regarding the question of additional light states, while the absence of additional singularities in the prepotential is a suggestive criterion for the absence of such states in a given region, it is not necessarily fully reliable. E.g., consider a family of Calabi-Yau manifolds $\cX$ and a point $z$ in moduli space at which the lattice
\begin{equation} \label{eq:intersection_with_H21}
    \left(H^{2,1}(\cX_{z}) \oplus H^{1,2}(\cX_{z})\right) \cap H^{3}(\cX_z,\IZ)
\end{equation}
is at least of rank 2. At such a point, infinitely many non-proportional D-brane charges lead to vanishing central charge, yielding an infinite number of candidates for massless states. Points on moduli space satisfying this constraint on the cohomology lattice of the associated manifold exist, and are indeed very special. As we discuss in appendix \ref{appendix:SUSY_vacua}, if the lattice \eqref{eq:intersection_with_H21} has a rank 2 sublattice whose complexification has a Hodge decomposition, such points coincide with supersymmetric vacua of type IIB flux compactifications. We leave the investigation of the intriguing question of additional massless states at such points for future study. 

In the remaining part of this section, we make some preliminary remarks regarding the choice of a distinguished symplectic frame away from large radius, and the relation to time reversal invariance. At a generic point on moduli space, such a choice will not exist. At conifold points, a family of distinguished frames is well-motivated in \cite{Huang:2006hq} in the context of imposing the so-called gap condition on topological string amplitudes. In the same reference, a tentative criterion is also put forth for orbifold points. Here, we would like to make an observation regarding a more subtle class of distinguished points on moduli space, so-called attractor points of rank 2, defined by the condition that the lattice
\begin{equation}
    H^{3}(\cX_z,\IZ) \cap \left(H^{3,0}(\cX_{z}) \oplus H^{0,3}(\cX_{z})\right) 
\end{equation}
have rank 2. As we discuss in appendix \ref{appendix:SUSY_vacua}, these points coincide with the supersymmetric vacua discussed above in one-parameter models.

Our simple observation is the following: at such points, the gauge coupling matrix $\cN$ can be put in block diagonal form via a rational symplectic transformation.
Identifying the graviphoton and its magnetic dual with the modes of $C_4$ (recall that we are considering type IIB compactification)\footnote{As the back and forth between IIA, IIB, the compactification manifold $X$, and its mirror $\check X$ can be mind-bending, let us restate the situation: we are discussing the gauge coupling matrix $\cN$ as obtained at a distinguished point on the K\"ahler moduli space of $X$ upon type IIA compactification. We perform our computation by considering the mirror image $z$ of the point, which is a distinguished point on the complex structure moduli space of $\check X$, and obtain $\cN$ via type IIB compactification on $\check X$.} of Hodge type $(3,0) \oplus (0,3)$ \cite{Billo:1995ge}, one block of unit size determines the gauge coupling and theta angle for the graviphoton, and the other of size $b_2(\cX_\fatdot) \times b_2(\cX_\fatdot)$ determines the couplings for the remaining vector fields. This form of $\cN$ would indicate the decoupling of the graviphoton from the vector multiplets at rank 2 attractor points.

To argue for the form of $\cN$, we first introduce the two lattices
\begin{eqnarray}
    \Lambda = H^3(\cX_z,\IZ) \cap \left(H^{3,0}(\cX_z) \oplus H^{0,3}(\cX_z) \right)\,, \label{eq:Lambda}\\
    \Lambda^\perp = H^3(\cX_z,\IZ) \cap \left(H^{2,1}(\cX_z) \oplus H^{1,2}(\cX_z) \right)\,,
\end{eqnarray}
as well as the notation $\Lambda_{\IQ} = \Lambda \otimes \IQ$ and $\Lambda^\perp_{\IQ} = \Lambda^\perp \otimes \IQ$. When $\Lambda$ is of rank 2, we can choose two elements $\alpha^0, \beta_0$ of a symplectic basis satisfying \eqref{eq:dual_basis} and underlying both the mode expansion of $C_4$ and the definition of the period vector associated to $\Omega$ (see \eqref{eq:Omega_expansion}) to lie in $\Lambda_{\IQ}$, and the remaining basis elements to lie in $\Lambda^\perp_\IQ$. With this choice, $X^i = F_i = 0$ and $\nabla_i X^0 = \nabla_i F_0 = 0$ for $i\neq 0$. We can immediately conclude from the presentation \eqref{eq:sol_N} of the gauge coupling matrix that with this choice of representatives of a basis of $H^3(X, \IQ)$,
\begin{equation} \label{eq:diagonal_N}
 \cN_{\IQ} =   \begin{pmatrix}
    F_0/X^0 & 0 \\
    0 & \overline{\nabla_\fatdot F_\fatdot}(\overline{\nabla_\fatdot X^\fatdot})^{-1}
    \end{pmatrix} \,,
\end{equation}
with $\nabla_\fatdot F_\fatdot$ and $\nabla_\fatdot X^\fatdot$ denoting the matrices with entries $\nabla_i F_j$ and $\nabla_i X^j$ respectively. 

The astute reader has undoubtedly remarked the unsettling multiple appearance of the field $\IQ$ in the preceding two paragraphs. The necessity of tensoring by $\IQ$ arises because the lattice $\Lambda \oplus \Lambda^\perp$ is generically of finite index in $H^3(\cX_z,\IZ)$. To permit the normalization $\int \alpha_I \wedge \beta^I = 1$, we will hence generically require recourse to a rational normalization of elements in $\Lambda$. Put differently, to reach the form \eqref{eq:diagonal_N} from a properly normalized symplectic basis for $H^3(\cX_z,\IZ)$ requires acting with an element of the rational symplectic group $\mathrm{Sp}(2(b_2+1),\IQ)$. This is troubling because, as outlined in the opening paragraph of appendix \ref{app:sympl_vs_monodromy}, the same symplectic transformation that acts on the period vector also acts on the vector $(G^-, F^-)$ of field strengths. A rational symplectic transformation of the form
\begin{equation}
    S = \frac{1}{r} \tilde{S} \,, \quad \tilde{S} \in \text{Mat}_{n \times n}(\IZ)
\end{equation}
with $n=2(b_2+1)$ and $r \in \IN$ chosen minimally would be permissible in the case of a charge lattice only populated in $r$ multiples of the elementary charges. We leave the investigation of this question, together with the more pressing conundrum regarding the spectrum of light states at rank 2 attractor points, for future study. 

If a legitimate symplectic frame in which $\cN$ takes the form \eqref{eq:diagonal_N} exists, it is tempting, particularly in the one-parameter case, to explicitly compute the diagonal entries of $\cN$ to see whether they are in any way distinguished. In the context of this paper, a natural question is whether they yield $\theta$ angles that are either $0$ or $\pi$, i.e.\ that preserve time reversal invariance without recourse to a monodromy symmetry. This turns out not to be the case. As we feel that the computation of these diagonal entries itself is interesting, we will discuss one example despite this negative result.

Consider the family of Calabi-Yau manifolds associated to the Picard-Fuchs equation AESZ 34 \cite{almkvist2005} which has an attractor point of rank two at $z=-\frac{1}{7}$ \cite{Candelas:2019llw}.\footnote{There are two Calabi-Yau threefolds described in \cite{almkvist2005}; one of which is a free $\mathbb{Z}/10\mathbb{Z}$ quotient of the other. Here, for simplicity, we will only consider the quotient manifold. This corresponds to the case $\kappa=1$ in the notation of \cite{Candelas:2019llw}.} This Picard-Fuchs equation has the Riemann symbol 

\begin{equation}
\label{eq:RiemannSymbolOfAESZ34}
   \mathcal{P} \left\{~\begin{matrix}
 0 & \frac{1}{25} & \frac{1}{9} & 1  &\infty\\ \hline
~0 & 0 & 0 & 0 & 1 \\
~0 & 1 & 1 & 1 & 1 \\
~0 & 1 & 1 & 1 & 2\\
~0 & 2 & 2 & 2 & 2
\end{matrix}~z\right\}
\end{equation}

and it describes the variation of Hodge structure of a family of Calabi-Yau manifolds with Hodge number $h^{2,1}=1$. Mirror to this family is another family of Calabi-Yau manifolds with triple intersection number $D^3=12$, second Chern class $c_2\cdot D = 12$ and Euler characteristic $\chi= -8$. With this topological data in hand, we may compute the periods in an integral symplectic basis around the MUM point at $z=0$ (see appendix~\ref{app:specialkaehler}, in particular equation \eqref{eq:ChangeOfBasisMatrixAroundMUMPT}). By analytically continuing these solutions to the attractor point at $z=-\frac{1}{7}$, it was found numerically in \cite{Candelas:2019llw} that the periods $\Pi$ in an integral symplectic basis are given by
\begin{equation}
    \Pi(-\tfrac{1}{7})=\omega_1\begin{pmatrix}8 \\ -30 \\ 0 \\ 5 \end{pmatrix}+i\, \omega_2 \begin{pmatrix} 0 \\ 0 \\ 2 \\ 1 \end{pmatrix}\,,
\end{equation}
where\footnote{Note that, in comparison with the normalization in \cite{Candelas:2019llw}, our periods contain an additional factor of $(2\pi i)^3$ so that $\Omega$ is an algebraic form defined over $\mathbb{Q}$.}
\begin{equation}
    \omega_1=13.323239482723603\cdots~,~~~~\omega_2=-80.866444656616459\cdots\,.
\end{equation}
This implies that in terms of the  basis dual to the basis of $H_3(X,\IZ)$ with regard to which  the period vector $\Pi$ is expressed, a set of generators of $\Lambda$ is given by
\begin{equation} \label{eq:gens_Lambda}
    (4,-15,-5,0)^T~~~~~\text{and}~~~~~(0,0,2,1)^T \,.
\end{equation}
Similarly,
\begin{equation}
\label{eq:covderiv at attractor point}
    \nabla_z\Pi\left(-\tfrac{1}{7}\right) = \widetilde{\omega}_1  \begin{pmatrix} 3 \\ -6 \\ 0 \\ 1 \end{pmatrix}
    +i\,\widetilde{\omega}_2
    \begin{pmatrix} -7  \\ 14 \\ -10 \\ -5 \end{pmatrix} \,,
\end{equation}
where 
\begin{equation}
    \widetilde{\omega}_1=51.010880877055569\cdots~,~~~~\widetilde{\omega}_2 = -38.125487167252326\cdots~.
\end{equation}
Hence, with regard to the same basis as above, a set of generators of $\Lambda^\perp$ is given by
\begin{equation} \label{eq:gens_Lambda_perp}
    (3,-6,0,1)^T~~~~~\text{and}~~~~~(1,-2,-5,-1)^T~.
\end{equation}

One readily checks that the integral of the wedge product of the generators of $\Lambda$ is equal to $\pm7$ and the same is true for the generators of $\Lambda^\perp$. Furthermore, the integral of the wedge product of an element of $\Lambda$ with an element of $\Lambda^\perp$ is indeed zero, as follows from considerations of Hodge type. Thus, we find that $\Lambda\oplus\Lambda^\perp$ is an index $7^2$ sublattice of $H^3(\cX_z,\mathbb{Z})$. It is, therefore, impossible to assemble an integral symplectic basis of the third cohomology from elements of this lattice. However, normalizing the first period in \eqref{eq:gens_Lambda} and the second period in \eqref{eq:gens_Lambda_perp} by a factor of $-\frac{1}{7}$ yields a rational symplectic basis of this space which respects the Hodge splitting. The period vector $\Pi$ of the holomorphic three-form at $z=-\frac{1}{7}$ in terms of the standard basis at the MUM points is expressed in terms of this basis by multiplication by the matrix $S$,
\begin{equation}
    S = \frac{1}{7}\begin{pmatrix} 
    14 & 7 & 0 & 0 \\
    5 & 1 & 1 & -2\\
    -5 & 0 & -4 & 15\\
    0 & -7 & 21 & -42
    \end{pmatrix} \in \frac{1}{7}\mathrm{Sp}(4,\mathbb{Z})~.
\end{equation}
As discussed above, the new basis diagonalizes the coupling matrix, yielding
\begin{equation}
    \mathcal{N}_\mathbb{Q} = \begin{pmatrix}
    \frac{-7}{\tau_\Lambda+3} & 0 \\
    0 & \frac{3\tau_{\Lambda^\perp}-2}{14\tau_{\Lambda^\perp}-7}
    \end{pmatrix}\,,
\end{equation}
where
\begin{equation}
    \tau_\Lambda =-\frac{1}{2}+i\, 3.034789127729667\cdots~,~~~~\tau_{\Lambda^\perp} =\frac{1}{2} + i\, 0.373699556954729\cdots~.
\end{equation}
The numbers $\tau_\Lambda$ and $\tau_{\Lambda^\perp}$ are known to be the ratios of periods of modular forms of weight $4$ and $2$ respectively for the congruence subgroup $\Gamma_0(14)\subset \mathrm{SL}(2,\mathbb{Z})$ \cite{Candelas:2019llw}. For example, the number $\tau_{\Lambda^\perp}$ may be understood as the complex structure of the elliptic curve
\begin{equation}
    y^2+xy+y=x^3+4x-6
\end{equation}
with $j$-invariant equal to $\left(\frac{215}{28}\right)^3$. However, as emphasized in \cite{Kachru:2020abh}, this is only one among many possible rational models. The question of which (if any) rational models are singled out by string theory remains an open question.

\section{\boldmath{$CP$} symmetry} \label{sec:CP}
Parity symmetry $P$ is more subtle to define than time-reversal symmetry $T$ in the context of compactifications, as we expect $P$ to also act on the compactification manifold. In addition to the considerations of section \ref{sec:time_reversal}, we hence need to generalize the spatial involution $x^i \mapsto -x^i$ to the case of manifolds which generically do not permit global coordinates. We will consider this generalization in the next subsection, before turning to the ensuing action in four dimensions.

\subsection[The action of $CP$ on the internal manifold]{The action of \boldmath{$CP$} on the internal manifold} \label{subsec:CP_internal}
In \cite{Strominger:1985it}, Strominger and Witten note that composing 4d parity with an orientation reversing involutive isometry in the internal dimensions yields an orientation preserving map and is a symmetry of type I supergravity, whereas 10d parity acting on type I supergravity on $\IR^{1,9}$ is not. For a generic compactification manifold $X$, it is not clear whether such a map exists. When a family of compactification manifolds $\cX$ is constructed by considering hypersurfaces or complete intersections in complex projective space (or more generally, a product of weighted projective spaces), an orientation reversing involution can be constructed for those members of the family for which the coefficients of the defining equations lie in $\IR$. In the following, we will call this the real slice of complex structure moduli space. If we refer to the $\IC$-valued solution set of these equations as $\cX_z(\IC)$, with $z$ specifying one such choice of real coefficients, complex conjugation defines an involution $c$:
\begin{eqnarray}
    c: \cX_z(\IC) &\rightarrow& \cX_z(\IC) \nn \\
    p &\mapsto& \bar{p} \,.
\end{eqnarray}
When $\cX$ describes a family of Calabi-Yau varieties with $h^{1,1} = 1$, $c$ induces an isometry of the Ricci flat metric $g$ on $\cX_z$ for any choice of K\"ahler class $[\omega]$. To see this, note that $c^*g$ is also Ricci flat. We thus need to show that the associated K\"ahler class is equal to $[\omega]$; Yau's theorem will then allow us to conclude that $g = c^*g$. We can compute this class as follows:
\begin{equation}
    c^* [\omega] = \alpha [\omega] 
\end{equation}
as $h^{1,1} = 1$ , and $\alpha = \pm 1$ by $c^2 =1$. As $c$ is orientation reversing,
\begin{equation}
\int c^*(\omega \wedge \omega \wedge \omega) = - \int \omega \wedge \omega \wedge \omega \,,
\end{equation}
hence $\alpha = -1$. Note that $-c^* \omega$ is the K\"ahler form associated to the metric $c^* g$, as $c$ maps the complex structure $J$ of $\cX_z$ to its negative,
\begin{equation} \label{eq:pullback_Kaehler_form}
    -c^* \omega = -c^* ( g \circ J \otimes 1) = c^*(g) \circ J \otimes 1 \,,
\end{equation}
thus concluding the demonstration. For $h^{1,1}>1$, by \eqref{eq:pullback_Kaehler_form}, we still require $[\omega] = - c^* [\omega]$ to conclude $c^*g = g$. However, we now must restrict the K\"ahler classes we consider to the subset satisfying this constraint. This subset is not empty: given any metric $g$ on $\cX_z$ with associated K\"ahler form $\omega$, $g + c^*g$ again defines a metric, with associated K\"ahler form $\omega - c^* \omega$ satisfying the constraint. 

In the following, we will restrict to the region in the the complex structure moduli space and the K\"ahler cone for which $c$ exists and describes an isometry.

To work out the effect of this involutive isometry on the compactified theory, we need to compute the pullback $c^*$ of the involution on representatives of the cohomology of $\cX_z(\IC)$ that enter into the compactification. Assuming that $\cX_z$ is a Calabi-Yau manifold, we can argue as follows for the middle dimensional cohomology: the space $H^0(\cX_z, \Omega^3)$ of global sections of the sheaf of algebraic 3-forms $\Omega^3$ is one-dimensional. A choice of section (obtained via a residue formula on the ambient space) exists~\cite{MR229641}  
which does not involve complex coefficients. Calling this choice $\Omega$, we thus have
\begin{equation} \label{eq:c_star_Omega}
    c^* \Omega = \bar{\Omega} \,.
\end{equation}
The authors of \cite{Strominger:1985it} observe that this transformation implies that in heterotic compactifications, charged matter in a representation $R$ of the gauge group is mapped to the conjugate representation $\bar{R}$. They thus identify the combined action of parity $\cP_4$ in 4d and $c$ on the compactification manifold with the discrete symmetry $\cC \cP$. Type II compactifications on smooth Calabi-Yau manifolds do not give rise to charged matter, but it is natural to identify this action with $\cCP$ also in this case, and we will do so:
\begin{equation} \label{eq:def_CP}
    \cCP = \cP_4 \circ c \,.
\end{equation}

To work out the action of $c^*$ on a basis $\{ \gamma_i \, | \, 1 \le i \le b_3 \}$ of $H^3(\cX_z,\IZ)$,  we write
\begin{equation}
    \Omega = \sum \gamma_i  \Pi_i \quad \Rightarrow \quad c^* \Omega = \sum \gamma_i \overline{\Pi}_i  = \sum (c^* \gamma_i) \, \Pi_i \,.
\end{equation}
The first equality on the RHS follows by \eqref{eq:c_star_Omega} and reality of $\gamma_i$, and the second by linearity of the pullback map. To solve for the $b_3$ forms $c^* \gamma_i$, we need $b_3$ linearly independent equations of this form. These can be obtained by applying the same reasoning to a full algebraic basis of $H^3(\cX_z)$. Concretely, such a basis can be constructed by considering derivatives of $\Omega$ with regard to the $\frac{b_3-2}{2}$ coordinates $z_i$ on complex structure moduli space, the latter obtained as rational functions of the coefficients of the defining equations. In the one-parameter case, e.g., away from apparent singularities, it suffices to consider $\Omega$ together with its derivatives up to and including order three. Writing
\begin{equation} \label{eq:def_W}
    (\Omega, \Omega' , \Omega'', \Omega''')_i = \sum_j \gamma_j \cW_{ji} \,,
\end{equation}
we obtain 
\begin{equation} \label{eq:c_star_from_W}
    \sum_j \gamma_j \overline{\cW}_{ji} = \sum_j c^*(\gamma_j) \cW_{ji} \quad \Rightarrow \quad
    c^* \gamma_i = \sum_j \gamma_j (\overline{\cW} \cdot \cW^{-1})_{ji} \,. 
\end{equation}
Note that as $c^*$ maps $H^3(\cX_z,\IZ)$ to $H^3(\cX_z,\IZ)$, the entries of the associated matrix are necessarily integers. Furthermore, on any connected subset of the real slice of moduli space that we are considering, $c$ is continuous. Hence, the matrix associated to $c^*$ is constant on such sets. We remark further that the determinant of $c^*$ (and more generally that of the pullback of any orientation reversing diffeomorphism) acting on the middle cohomology equals $(-1)^{\frac{b_3}{2}}$. This can be seen as follows: the vector space $H^3(\cX_z,\mathbb{C})$ together with the pairing ${(\omega_1,\omega_2) \mapsto \int_{\cX_z} \omega_1 \wedge \omega_2}$ is a symplectic vector space. We have $\int_{\cX_z} c^* \omega_1 \wedge c^* \omega_2 = \int_{\cX_z} c^* (\omega_1 \wedge \omega_2)$; since $c$ is orientation reversing (because $\cX_z$ is a threefold), this shows that $c^*$ is antisymplectic. Hence, $ic^*$ is symplectic and thus has determinant 1. The claim follows by $\det(c^*) = (-i)^{b_3} \det(ic^*)$.

Turning next to even dimensional cohomology, $H^2(\cX_z)$ and $H^4(\cX_z)$ can be decomposed into an even and an odd eigenspace of $c^*$. Furthermore, integration gives a non-degenerate pairing $H^2(\cX_z,\mathbb{R}) \otimes H^4(\cX_z,\mathbb{R}) \rightarrow \mathbb{R}$ which is anti-invariant under $c^*$ since $c$ is orientation reversing. Hence, the pairing of forms of equal parity must vanish, allowing us in particular to conclude that there are (non-canonical) isomorphisms between the even eigenspace of $H^2(\cX_z)$ and the odd eigenspace of $H^4(\cX_z)$, and vice versa. Note that when we fix a K\"ahler form $\omega$ which is odd under $c^*$, such an isomorphism is induced canonically, as follows from the Lefschetz theorem, by wedging with $\omega$.

\subsection[$CP$ action in 10d and 4d]{\boldmath{$CP$} action in 10d and 4d} \label{subsec:CP_10d_action}
We will be interested below in flux compactifications. These take a simpler form in type IIB supergravity. The action of type IIB, ignoring the issue of the self-duality of the 5-form field strength, is given by
\begin{eqnarray}
    S^{\mathrm{IIB}} &=& \int \Big[ e^{-2\phi} \left( \frac{1}{2} R *1 + 2 \dd \phi \wedge * \dd \phi - \frac{1}{4} H_3 \wedge * H_3 \right) \label{eq:IIB_action}\\
    && - \frac{1}{2}  \left( F_1 \wedge * F_1 + \tF_3 \wedge * \tF_3 + \frac{1}{2} \tF_5 \wedge * \tF_5 \right) \nonumber\\
    && - \frac{1}{2}  C_4 \wedge H_3 \wedge F_3 \Big] \,, \nonumber
\end{eqnarray}
where $F_i = \dd C_i$ and
\begin{equation} \label{eq:def_tF}
    \tF_3 = F_3 - C_0 \wedge H_3 \,, \quad \tF_5 = F_5 - \frac{1}{2} C_2 \wedge H_3 + \frac{1}{2} B_2 \wedge F_3 \,.
\end{equation}
This action gives rise to the correct equations of motion, which must then be supplemented with the self-duality constraint $\tF_5 = * \tF_5$.

$\cCP$ as defined in equation \eqref{eq:def_CP} is an orientation preserving isometry; it is hence a symmetry of the action \eqref{eq:IIB_action} by the discussion in section \ref{subsec:trans_p_forms}, without the need of introducing intrinsic phases. The $\cCP$ invariance of the 4d theory arising upon compactification is thus automatic. Instantons correct the tree-level prepotential governing the vector multiplet sector. As the corrected prepotential is expressed via \eqref{eq:prepotential_from_periods} entirely in terms of the periods of $\Omega$, $c$ merely maps the prepotential to its complex conjugate, preserving the action.

While it is thus not necessary to assign intrinsic phases to 10d fields in order to preserve $CP$ invariance, we will see below that the freedom to do so extends the set of vacua which preserve $CP$ invariance. The phases we will require in our study of flux vacua in section \ref{subsec:CP_flux_vacua} coincide with those we must introduce in order for a space-filling D3 brane to preserve $CP$ symmetry: the WZW coupling \eqref{eq:WZW_coupling} of a space-filling D3 brane located at a fixed point of the $c$-action will be invariant under $CP$ only if we impose an intrinsic phase $-1$ for $C_0$. By \eqref{eq:def_tF}, this in turn requires that $C_4$ and either $B_2$ or $C_2$ also acquire an intrinsic phase $-1$. Either choice is consistent with the invariance of the topological term in \eqref{eq:IIB_action}. 

Note that phenomenological type IIB models often include space-filling D7 branes as ingredients, and invoke Euclidean D3 brane instantons to generate potentials for axions. Whether these  branes preserve or violate $CP$ symmetry depends on the action of $c$ on the internal cycle that they wrap.

\subsection[Spontaneously breaking $CP$ invariance]{Spontaneously breaking \boldmath{$CP$} invariance} \label{subsec:spontaneously_breaking_CP}

The transformation properties of the 4d fields in \eqref{eq:4d_action} under $CP$ depend on the intrinsic phase of their parent field in 10d as well as on the action of $c^*$ on the basis of forms on which the Kaluza-Klein reduction of the parent field is based.

We first consider the hypermultiplet sector: the reduction here is based on a basis of even cohomology. As argued above, this basis can be chosen with definite parity with regard to $c^*$. Depending on the choice of intrinsic phase for $C_0$, $C_2$, $C_4$ and $B_2$, it is the even or odd modes which transform trivially under $CP$ and whose VEV is thus compatible with $CP$ invariance. The map between these modes and the hyperscalars $\xi^A$ and $\txi_A$ occurring in the 4d action \eqref{eq:4d_action} is somewhat intricate; it is worked out in \cite{Bohm:1999uk}.

We turn next to the vector multiplet sector. The scalars here are functions of the complex structure moduli. As the definition of $c$ required restricting to a real slice of complex structure moduli space invariant under the action of $c$, we conclude that any function of these moduli is also invariant. Hence, any VEV within the real slice that the vector scalars take is compatible with $CP$ invariance. Note in particular that on this slice, all $\theta$ angles vanish.

In the following, we will focus on points on complex structure moduli space that correspond to supersymmetric flux vacua of type IIB string theory, as reviewed in appendix \ref{appendix:SUSY_vacua}. As we discuss there, these coincide with rank 2 attractor points in one-parameter models. We review a strategy to find such rank 2 attractor points in appendix \ref{appendix:Finding rank 2 attractors}. All of the examples listed in table \ref{tab:attractor} of this appendix indeed lie on the real slice of moduli space; the corresponding VEVs of the vector scalars are therefore invariant under the $CP$ transformation. This property however is not generic. In the final paragraph of appendix \ref{appendix:Finding rank 2 attractors}, we also give several rank 2 attractor points which do not lie on the real slice.

\subsection[$CP$ invariance of supersymmetric flux vacua]{\boldmath{$CP$} invariance of supersymmetric flux vacua} \label{subsec:CP_flux_vacua}

In the previous section, we discussed the $CP$ invariance of distinguished points on moduli space. To localize the theory at these points requires additional ingredients. In this section, we want to study the $CP$ invariance of the theory in the presence of non-trivial fluxes. Introducing such non-trivial backgrounds will break any symmetry which does not act trivially on these.

Compactifying IIB string theory on a family of Calabi-Yau threefolds $\cX$ with non-trivial $F_3$ and $H_3$ flux will generate a superpotential for some of the moduli (see appendix~\ref{appendix:SUSY_vacua}). More precisely, for given locally constant cycles $F_3,H_3\in H^3(\cX_{\boldsymbol{\cdot}},\mathbb{Z})$\footnote{In fact, the correct quantization condition is given in equation \eqref{eq:flux_quantization}. We will implicitly choose a normalization of the field strengths to absorb the factor of $l_s^2$ in this section so as to not overload the notation.} defined over a contractible subset of the moduli space, the superpotential is given by
\begin{equation}
    W(z) = \int_{\cX_z} G_3\wedge \Omega_z
\end{equation}
where
\begin{equation} \label{eq:def_g3}
    G_3=F_3-\tau H_3.
\end{equation}
Here, $\tau$ is a complex number that is identified with the vacuum expectation value of the axio-dilaton 
\begin{equation}
    \tau = C_0+ie^{-\phi}~.
\end{equation}
The superpotential $W$ depends on the complex structure moduli $z$ through the section $\Omega$ of the bundle of holomorphic (3,0)-forms. 

Traditionally, one specifies the theory by fixing $F_3$ and $H_3$ and then solving the ensuing 4d equations of motion to determine the vacuum expectation value for the complex structure moduli and the axio-dilaton which follow. As explained in appendix~\ref{appendix:SUSY_vacua}, the supersymmetric solutions to these equations determine a point $z=z_0$ in complex structure moduli space at which a rank 2 lattice $\Gamma \subset H^3(\cX_{z_0},\IZ)$ exists whose complexification has a Hodge decomposition of type $(2,1) \oplus (1,2)$. 

In this section, we consider such vacua from a slightly different vantage point: we ask whether given such a point $z_0$, it is possible to choose compatible fluxes $F_3$ and $H_3$ such that $CP$ is conserved. In particular, this requires the fluxes to be invariant under this transformation.

Let us assume that at $z_0$, the intersection
\begin{equation}
    \left(H^{2,1}(\cX_{z_0}) \oplus H^{1,2}(\cX_{z_0})\right) \cap H^3(\cX_{z_0},\IZ)
    \label{eq:Intersection2dim}
\end{equation}
has exactly rank 2, and hence equals $\Gamma$ (we will comment on higher rank intersections at the end of this section). The intersection is clearly invariant by $c^*$: $c^*$ acts as an involution on $H^3(\cX_{z_0},\IQ)$ as $c: \cX_{z_0} \rightarrow \cX_{z_0}$ is an involution. Furthermore, as $c$ is antiholomorphic, $c^* : H^{2,1}(\cX_{z_0}) \xrightarrow{\,\smash{\raisebox{-0.55ex}{\ensuremath{\scriptstyle\sim}}}\,} H^{1,2}(\cX_{z_0})$. As both $H^3(\cX_{z_0},\IQ)$ and $H^{2,1}(\cX_{z_0}) \oplus H^{1,2}(\cX_{z_0})$ are invariant under $c^*$, so is their intersection. 

As an involution, $c^*$ restricted to $\Gamma$ is diagonalizable over $\IQ$ with eigenvalues $\pm 1$. 
The argument at the end of section \ref{subsec:CP_internal} then shows that $\Gamma_{\IQ}$ decomposes into a sum of one-dimensional eigenspaces $E_{\pm}$ to eigenvalue $\pm 1$ respectively. We shall call the indivisible integral eigenvectors in these eigenspaces $\gamma_+$ and $\gamma_-$. 

$CP$ conservation now faces an apparent quandary: due to the constraint \eqref{eq:crossed_fluxes}, we cannot choose both $F_3$ and $H_3$ as multiples of the invariant form $\gamma_+$. The conclusion that no supersymmetric vacuum preserves $CP$ would however be too hasty, as happily, we have the freedom to introduce additional intrinsic phases, as discussed in section \ref{subsec:CP_10d_action} above. 

Choosing an intrinsic $CP$ phase $-1$ for either $F_3$ or $H_3$, for the flux background to not break $CP$, the field strength carrying this phase should be an integer multiple of $\gamma_-$, the field strength carrying the phase $+1$ an integer multiple of $\gamma_+$. Following the reasoning around \eqref{eq:4_choices_to_identify_G3} in the appendix, we conclude that both choices $F_3 \in E_+$, $H_3 \in E_-$ and $H_3 \in E_+$, $F_3 \in E_-$ are possible to fix the vacuum to the point $z_0$. Whether this vacuum is $CP$ symmetric hence depends only on the value of $C_0$: recall that when imposing a non-trivial intrinsic phase on either $H_3$ or $F_3$, $C_0$ also acquires an intrinsic $CP$ phase -1. Its VEV must therefore vanish in order to preserve $CP$ invariance. One may be tempted to enlarge the $CP$ preserving domain by virtue of the discrete shift invariance subgroup of S-duality. This is not possible, as this invariance is fixed by requiring that $G_3$ have the form 
\begin{equation} \label{eq:decomp_G3}
    G_3 = \gamma_+ - \tau \gamma_- \quad \text{or} \quad G_3 = \gamma_- - \tau \gamma_+ \,.
\end{equation}
For multi-parameter models, a computation must determine whether given $\Gamma$, the decomposition \eqref{eq:decomp_G3} occurs with $\reP \tau =0$ or not. If so, the vacuum is $CP$ preserving, else $CP$ violating. Specializing however to one-parameter models, we can show that all supersymmetric vacua are $CP$ preserving: we define a distinguished generator $\gamma$ of $\Gamma_{\IC}$ in this case as
\begin{equation}
    \gamma = \nabla_z \Omega = (\partial_z + K_z)\Omega \,,
\end{equation}
with $\Omega$ chosen to satisfy \eqref{eq:c_star_Omega}. As $K \in \IR$, we also have $K_z \in \IR$ in the slice of moduli space under consideration. Hence,
\begin{equation}
    c^* \gamma = \bar{\gamma} \,.
\end{equation}
Therefore, $\reP \gamma \in (E_+)_{\IC}$, $\imP \gamma \in (E_-)_{\IC}$, i.e.
\begin{equation}
    \gamma = \alpha \gamma_+ + i\, \beta \gamma_- \quad \text{with} \quad \alpha, \beta \in \IR \,,
\end{equation}
allowing us to conclude that all choices of $G_3$ compatible with this flux vacuum have $C_0 = 0$, hence preserve $CP$.

More generally, in the case of multi-parameter models and even if the intersection \eqref{eq:Intersection2dim} has rank greater than 2, this conclusion can be made if e.g.\ $\Gamma$ is cut out by a correspondence defined over $\mathbb{R}$. Then $\Gamma_\mathbb{C}$ can be generated by forms from $H^3(\cX_{z_0},\IZ)$, which are invariant under $\gamma \mapsto \overline{\gamma}$, but also by forms from the algebraic de Rham cohomology $H^3_{\text{dR}}(\cX_{z_0})$, on which $c^*$ acts by $\gamma \mapsto \overline{\gamma}$. For more details on correspondences we refer to \cite{Motives}. 

As an example, consider again the Picard-Fuchs equation AESZ 34 (see equation~\eqref{eq:RiemannSymbolOfAESZ34} for its Riemann symbol) and the associated family of Calabi-Yau manifolds described above. This family has an attractor point of rank two at $z=-\frac{1}{7}$ where the Hodge structure splits as in \eqref{eq:HodgeSplitting}. The computation of the pullback $c^*$ of the complex conjugation map at this point is simplified by the fact that this family exhibits no singularities on the negative real axis of moduli space. By our discussion below equation \eqref{eq:c_star_from_W}, we can hence evaluate the matrix $\cW$ introduced in \eqref{eq:def_W} immediately to the left of the MUM point. To this end, let 
\begin{equation}
    \Pi = \begin{pmatrix}
    F_I\\X^I
    \end{pmatrix}
\end{equation}
denote the period vector in the integral symplectic basis of the third cohomology adapted to the MUM point (see the discussion around \eqref{eq:period_at_MUM}), such that
\begin{equation}
    \mathcal{W} = (\Pi,\Pi',\Pi'',\Pi''') \,.
\end{equation}
Identifying $c^*$ with its matrix expression in this basis, we must thus evaluate
\begin{equation}
\label{eq:pullbackofCCmap}
    c^* = \overline{\cW} \cdot \cW^{-1} \,.
\end{equation}

The period vector $\Pi$ can be obtained by applying a universal matrix $T$, depending only on the topological data of the mirror Calabi-Yau threefold, to the period vector $\varpi$  in the Frobenius basis at the MUM point (see equation~\eqref{eq:ChangeOfBasisMatrixAroundMUMPT}). As the coefficients of the holomorphic functions $g_i(z)$ on which $\varpi$ depends (see \eqref{eq:Frobenius}) are rational, the imaginary contributions to $\varpi$ evaluated on the negative real axis arise only from the evaluation of the logarithms occurring in this expression. Analytically continuing along the upper half plane, we can thus write
\begin{equation}
    \cW = T  
    \begin{pmatrix}
     1 & 0 & 0 & 0 \\
    \pi  i & 1 & 0 & 0 \\
    \frac{(\pi  i)^2}{2} & \pi  i & 1 & 0 \\
    \frac{(\pi  i)^3}{6} & \frac{(\pi  i)^2}{2} & \pi  i & 1 \\
    \end{pmatrix} \begin{pmatrix*}[r]
        g_0(z) \\
        g_0(z)\log(|z|) + g_1(z) \\
        \frac{1}{2} g_0(z) \log^2(|z|) +g_1(z) \log(|z|) + g_2(z) \\
        \frac{1}{6} g_0(z) \log^3(|z|) + \frac{1}{2}g_1(z) \log^2(|z|) + g_2(z) \log(|z|) + g_3(z) 
    \end{pmatrix*}
\end{equation}
for $z \in (-\infty,0)$. The final matrix in this expression, being real, does not contribute to \eqref{eq:pullbackofCCmap}, such that $c^*$ is  determined solely by topological data \cite{Yang:2020sfu,Yang:2020lhd}:
\begin{equation}
c^* =
    \begin{pmatrix}
    1 & 1 & -\frac{1}{12}(c_2\cdot D+2D^3) & \frac{D^3}{2}-\sigma\\
    0 & -1 & \frac{D^3}{2}-\sigma & -D^3+2\sigma\\
    0 & 0 & -1 & 0\\
    0 & 0 & -1 & 1
    \end{pmatrix} =
    \begin{pmatrix}
    1 & 1 & -3 & 6\\
    0 & -1 & 6 & -12\\
    0 & 0 & -1 & 0\\
    0 & 0 & -1 & 1
    \end{pmatrix} \,;
\end{equation}
we have used that for the example under consideration, $D^3=12$, $c_2 \cdot D = 12$  and ${\sigma=0}$.

Normalizing the covariant derivative \eqref{eq:covderiv at attractor point} appropriately, we can now check explicitly that it gives rise to a flux $G_3$, which when expressed in the same basis underlying the expressions \eqref{eq:gens_Lambda} and \eqref{eq:gens_Lambda_perp} is given by
\begin{equation}
G_3 = \begin{pmatrix} 3 \\ -6 \\ 0 \\ 1\end{pmatrix}
-\tau \begin{pmatrix} -7 \\ 14 \\ -10 \\ -5 \end{pmatrix} 
\end{equation}
with
\begin{equation}
    \tau = i \,0.747399113909459\cdots~.
\end{equation}
Hence, $c^* F_3 = F_3$, $c^* H_3 = - H_3$, and $C_0 =0$. It is thus indeed possible to choose $CP$ invariant fluxes which stabilize the theory to a $CP$ invariant supersymmetric vacuum at $C_0=0$ and $z =- \frac{1}{7}$.

\acknowledgments

We would like to thank Janis D\"ucker, Daniel Huybrechts, Dominic Joyce, Christian Kaiser, Spiro Karigiannis, Boris Pioline, Duco van Straten and Stefan Theisen for useful conversations.

K.B.\ is supported by the International Max Planck Research School on Moduli Spaces of the Max Planck Institute for Mathematics in Bonn. M.E.\ is supported by the US Department of Energy under grant DE-SC0010008. A.K.K.P. acknowledges support under ANR grant ANR-21-CE31-0021. A.K.\ likes to thank  Dr.\ Max R\"ossler,  the Walter Haefner Foundation and the  ETH Z\"urich Foundation for support.
\newpage

\appendix
\section{The gauge sector of 4d \boldmath{$\cN=2$} supergravity} \label{app:4d_SUGRA}
In this appendix, we review the structure of the gauge sector of 4d $\cN =2$ supergravity. We introduce the gauge coupling matrix $\bcN$ both as a linear map relating special geometry data and expressed in terms of a prepotential, and derive the equivalence of the two definitions when a prepotential exists. With regard to the action of the integral symplectic group in the gauge sector, we discuss the distinction between dualities and symmetries, and equate the latter to the action of the monodromy group when the supergravity is obtained via Calabi-Yau compactification.

\subsection{Vector multiplets and the gauge coupling matrix}

The vector multiplet sector of the 4d ${\cal N}=2$ effective action up to second order in derivatives is governed by special geometry. A sigma model with target space $\cM$ captures the dynamics of the scalars in this sector. Let $U \subset \cM$ be a contractible subset with complex coordinates $z_1,...,z_n$. In terms of holomorphic functions $\bX,\bF: U \rightarrow \mathbb{C}^{n+1}$, whose components we denote by $X^I$ and $F_I$ with $I=0,...,n$, the sigma model metric $g$ is determined by the K\"ahler potential
\begin{equation}
    K = - \log \left(2
    \imP \left( \bX^T \overline{\bF} \right) \right) 
\end{equation}
via
\begin{equation} \label{eq:sigma_model_metric}
    g_{i \bar{\jmath}} = \frac{\partial}{\partial z_i} \frac{\partial}{\partial \bar{z}_j} K \,.
\end{equation}

The kinetic and topological terms involving the vector fields are determined by the gauge coupling matrix $\boldsymbol{\cN} : U \rightarrow \mathbb{C}^{(n+1) \times (n+1)}$. It is defined by the relations
\begin{align}
    \bF &= \boldsymbol{\cN} \bX \label{eq:NIJ_bis1} \\
        \nabla_i \bF &= \overline{\boldsymbol{\cN}} \,\nabla_i \bX \quad \text{for} \quad i=1,...,n \, , \label{eq:NIJ_bis2}
\end{align}
where $\nabla_i = \partial_i + (\partial_i K)$. These equations can be solved to yield
\begin{equation}
{\cal N}= \Phi\,  \Xi^{-1}, \qquad {\rm with }\quad \Xi= (\bX \quad \overline{\nabla_1 \bX} \quad ... \quad \overline{\nabla_n \bX}), \ \    \Phi=(\bF \quad \overline{\nabla_{1} \bF} \quad ...\quad \overline{\nabla_{n} \bF})\, . 
\label{eq:sol_N}
\end{equation}
As $K$ is invariant under real symplectic transformations
\begin{equation} \label{eq:sympl_transf_FX}
    \begin{pmatrix}
    \bF \\
    \bX
    \end{pmatrix}
    \mapsto
    \begin{pmatrix}
    A & B \\
    C & D 
    \end{pmatrix}
    \begin{pmatrix}
    \bF \\
    \bX
    \end{pmatrix} \,,
\end{equation}
the associated transformation of the gauge coupling matrix 
\begin{equation} \label{eq:sympl_transf_N}
    \boldsymbol{\cN} \mapsto (A \boldsymbol{\cN} + B)(C \boldsymbol{\cN} + D)^{-1}
\end{equation}
follows immediately from \eqref{eq:sol_N}.

Both the positivity of the metric \eqref{eq:sigma_model_metric} and of the gauge kinetic term, which requires\footnote{Recall that the Maxwell action is $I = - \frac{1}{4} \int F \wedge *F$, with the sign ensuring that the energy density of the field is a positive multiple of $\boldsymbol{E}^2 + \boldsymbol{B}^2$.}
\begin{equation}  
{\rm Im}( \boldsymbol{\cN}) < 0 \,,
\label{eq:NIJ>0}
\end{equation} 
impose constraints on the functions $\bX$ and $\bF$. It is not possible to satisfy these constraints on all of target space $\cM$ using globally defined holomorphic functions. As we will review in the next section, the required structures emerge naturally when $\cM$ is identified with the moduli space of complex structures of a Calabi-Yau manifold.

We now consider the case that locally on $\cM$, we can introduce the coordinates $z_i = \frac{X^i}{X^0}$ as well as a homogeneous function $F$ of degree 2 in $X^0,...,X^n$, called a prepotential, such that $F_I = \partial_{X^I}F $. In particular, when $\cM$ coincides with the complex structure moduli space of a Calabi-Yau manifold, such a local choice is always possible. The gauge coupling matrix is then given by \cite{Ceresole:1995jg}
\begin{equation} 
\boldsymbol{\cN}=\overline{\boldsymbol{\mathcal{F}}} + 2 i \frac{ ({\rm Im}\boldsymbol{\mathcal{F}} \bX) ({\rm Im}\boldsymbol{\mathcal{F}} \bX)^T} { \bX^T{\rm Im} \boldsymbol{\mathcal{F}} \bX} \, ,  
\label{eq:NIJ} 
\end{equation}
where $\boldsymbol{\mathcal{F}} : U \rightarrow \mathbb{C}^{(n+1)\times (n+1)}$ is defined by $\boldsymbol{\mathcal{F}}_{IJ} = \partial_{X^I}\partial_{X^J} F$. To demonstrate this equality, we will show that this expression satisfies the relations \eqref{eq:NIJ_bis1} as well as the relation
\begin{align}
    (\partial_i \boldsymbol{\cN}) \bX = (\overline{\boldsymbol{\cN}}-\boldsymbol{\cN})\nabla_i \bX  \, ,
\end{align}
which is equivalent to \eqref{eq:NIJ_bis2} given \eqref{eq:NIJ_bis1}. For the length of this demonstration, let $\boldsymbol{\cN}$ indicate the expression \eqref{eq:NIJ}. By the homogeneity and symmetry of $\boldsymbol{\mathcal{F}}$,
\begin{align}
    \boldsymbol{\cN} \bX = \overline{\boldsymbol{\mathcal{F}}} \bX + 2i \, {\rm Im}\boldsymbol{\mathcal{F}} \bX = \boldsymbol{\mathcal{F}} \bX = \bF \,.
\end{align}
This is relation (\ref{eq:NIJ_bis1}). With\footnote{We use that a homogeneous function of degree 0 in $X^0,...,X^n$ can be seen as a function of $z_1,...,z_n$ and vice versa, and that with this identification one has $\partial_i = X^0 \partial_{X^i}$.} $\partial_i \boldsymbol{\cN} = X^0 \partial_{X^i}\boldsymbol{\cN}$ and 
\begin{equation}
\begin{aligned}
    \nabla_i \bX &= \partial_i \bX + (\partial_i K) \bX\\
    &= X^0\left(\partial_i \frac{\bX}{X^0}+(\partial_i (K+\log(|X^0|^2))) \frac{\bX}{X^0}\right)\\
    &= (X^0)^2\left(\partial_{X^i} \frac{\bX}{X^0}+(\partial_{X^i} (K+\log(|X^0|^2))) \frac{\bX}{X^0}\right) \\
    &= X^0 (\partial_{X^i} \bX + (\partial_{X^i} K) \bX)\, ,
\end{aligned}
\end{equation}
where we now treat $K$ as a function of $X^0, ..., X^n$, it only remains to show that 
\begin{align} \label{eq:step_N_prepot}
    (\partial_{X^i} \boldsymbol{\cN})\bX = (\overline{\boldsymbol{\cN}}-\boldsymbol{\cN})(\partial_{X^i}\bX + (\partial_{X^i}K) \bX) \, .
\end{align}
Using that $X^I\partial_{X^I}$ annihilates any homogeneous function of degree 0, we get
\begin{align}
    \partial_{X^i}K = -\partial_{X^i}\log \left(2
    \imP \left( \bX^T \overline{\bF} \right) \right) = -\partial_{X^i}\log \left(-2 \overline{\bX}^T \text{Im} \boldsymbol{\mathcal{F}} \bX     \right) = -\frac{(\text{Im} \boldsymbol{\mathcal{F}} \overline{\bX})_i}{\overline{\bX}^T \text{Im} \boldsymbol{\mathcal{F}} \bX} \,.
\end{align}
Thus, the $J^{\text{th}}$ component of the RHS of \eqref{eq:step_N_prepot} evaluates to
\begin{equation}
\begin{aligned}
(\overline{\boldsymbol{\cN}}-&\boldsymbol{\cN})_{Ji} + (\partial_{X^i}K) (\overline{\mathcal{N}} \bX - \bF)_J\\
=& \boldsymbol{\mathcal{F}}_{Ji} - \boldsymbol{\cN}_{Ji} -2i\left( \left(\frac{ ({\rm Im}\boldsymbol{\mathcal{F}} \overline{\bX}) ({\rm Im}\boldsymbol{\mathcal{F}} \overline{\bX})^T} { \overline{\bX}^T{\rm Im} \boldsymbol{\mathcal{F}} \overline{\bX}}\right)_{Ji}+(\partial_{X^i}K) \left( \frac{ ({\rm Im}\boldsymbol{\mathcal{F}} \overline{\bX}) ({\rm Im}\boldsymbol{\mathcal{F}} \overline{\bX})^T} { \overline{\bX}^T{\rm Im} \boldsymbol{\mathcal{F}} \overline{\bX}}\bX\right)_J \right) \\
=& \boldsymbol{\mathcal{F}}_{Ji} - \boldsymbol{\cN}_{Ji} \\
=& \partial_{X^i}(\boldsymbol{\cN}\bX)_J - (\boldsymbol{\cN} \partial_{X^i} \bX)_J \\
=& ((\partial_{X^i} \boldsymbol{\cN})\bX)_J \, .
\end{aligned}
\end{equation}
This concludes the demonstration.

\subsection{Symplectic transformations and monodromy symmetry} \label{app:sympl_vs_monodromy}
The Bianchi identities and equations of motion for the gauge field strengths $F$ can be compactly formulated in terms of the linear combination \cite{MR2920151} 
\begin{equation}
    F^- = F + i*F \,.
\end{equation}
The dual gauge field strengths
\begin{equation}
    G^- =  \overline{\bcN}F^-
\end{equation}
combine with the $F^-$ to form a vector $(G^-,F^-)^T$. Given the transformation behavior \eqref{eq:sympl_transf_N} of the gauge coupling matrix $\bcN$ under symplectic transformations, the transformation
\begin{equation} \label{eq:sympl_transf_F}
    F^- \rightarrow (F^-)'=(C G^- + D F^-) = (C \overline{\bcN} + D) F^-
\end{equation}
implies
\begin{equation}
    G^- \rightarrow (A \overline{\bcN} + B)(C \overline{\bcN} + D)^{-1} (F^-)' =  (A \overline{\bcN} + B) F^- = (A G^- + B F^-) \,,
\end{equation}
i.e.\ the vector $(G^-, F^-)^T$ transforms as a symplectic vector. A simultaneous symplectic transformation of the vectors $(\bF, \bX)^T$ and $(G^-, F^-)^T$ leads to an equivalent theory, in that the equations of motion are invariant under this transformation. The action is invariant for the subset of transformations for which $B=C=0$. The complement is sometimes referred to as the set of non-perturbative symplectic transformations.

Among all symplectic transformations, those count as {\it symmetries} (in contradistinction to {\it dualities}) which are induced by a transformation of the fundamental fields of the theory. As we have just seen, the transformation of the vector of field strengths $(G^-,F^-)$ is induced by the transformation \eqref{eq:sympl_transf_F} of the fundamental fields $F^-$, given the transformation \eqref{eq:sympl_transf_N} of the gauge coupling matrix $\bcN$. The latter in turn follows from the definition \eqref{eq:sol_N} upon the symplectic transformation \eqref{eq:sympl_transf_FX} of the vector $(\bF, \bX)^T$. The distinction between symmetry and duality hence boils down to the question whether \eqref{eq:sympl_transf_FX} can be induced by a transformation of the coordinates $z_i$ on moduli space. For a choice of vector $(\bF, \bX)^T$ for which $\bX$ can serve as local coordinates, a symplectic transformation thus qualifies as a symmetry if
\begin{equation}
    \bF(C \bF(\bX) + D \bX) = A \bF(\bX) + B \bX \,.  
\end{equation}
A transformation on the other hand which requires changing the functional form of $\bF$ can be interpreted as requiring a change of the coupling constants of the theory, and hence yields a different, though dual, theory.

Symplectic symmetries are encountered naturally in the context of Calabi-Yau compactifications, in which the vector $(\bF, \bX)^T$ is identified with the period vector of the holomorphic 3-form $\Omega$: they arise upon analytic continuation of the period along a closed path (which deserves to be interpreted as a transformation of $z_i$). The full group of such transformations is called the monodromy group of the Calabi-Yau manifold and can be identified with the symplectic symmetry group of the $\cN=2$ theory. 
We specialize to $\cN=2$ theories in the context of Calabi-Yau compactifications in the next section.

\section{Special geometry and the complex deformation space of Calabi-Yau manifolds}  
\label{app:specialkaehler}
We review the special geometry of the moduli space of complex structures of Calabi-Yau manifolds in this appendix. We pay special attention to the question of positivity of the K\"ahler metric and the gauge coupling matrix, as well as to the question of the integrality of the monodromy transformations around the MUM point.

\subsection{Introducing Calabi-Yau manifolds and mirror symmetry}
Different authors choose to include different amounts of data in their definition of Calabi-Yau manifolds, see \cite{MR3965409,MR1963559} for reviews. We will define an $n$-dimensional Calabi-Yau manifold as a compact Kähler manifold $(X,\omega)$, $\omega$ indicating the K\"ahler form, with holonomy group $\mathrm{SU}(n)$. An important consequence of this definition is that the Hodge numbers $h^{p,0}$ vanish for $p \neq 0,n$. For $n \geq 3$, this in particular implies $h^{2,0} = 0$; using the Kodaira embedding theorem, one can then show that $X$ is isomorphic to the vanishing locus of homogeneous polynomials in some $\mathbb{C}\mathbb{P}^N$. Another important consequence of the definition is that the canonical bundle of $X$ is trivial, i.e.\ there exists a nowhere vanishing holomorphic $(n,0)$-form $\Omega$ which is unique up to multiplication by complex numbers.\footnote{The uniqueness as a form rather than merely a cohomology class follows from the fact that a holomorphic $(n,0)$-form on a complex $n$-dimensional manifold is $\bar{\partial}$ and $\bar{\partial}^\dagger = -*\bar{\partial}*$ closed, hence harmonic.} The  space of infinitesimal complex structure deformations, of dimension  ${\rm dim}\, H^1(X,TX) = {\rm dim}\, H^{n-1,1}(X) = h^{n-1,1}$, is globally unobstructed~\cite{MR915841,MR1027500}, giving rise to a natural deformation family  $\cX \rightarrow \cM_{\text{cs}}$ 
over the complex structure moduli space $\cM_{\text{cs}}$.  
Choosing a holomorphic section $z \mapsto \Omega_z$ of nowhere vanishing holomorphic $(n,0)$-forms over $\mathcal{M}_{\text{cs}}$, one can define a global Kähler potential $K$ on $\mathcal{M}_{\text{cs}}$ by
\begin{equation} 
\label{eq:Kaehlerpotential}
e^{-K(z, \bar z)}= i^{n^2} \int_{\cX_z}\Omega_z \wedge \bar \Omega_z \, .
\end{equation} 
This gives $\mathcal{M}_{\text{cs}}$ the structure of a Kähler manifold; the associated metric is called the Weil-Petersson metric. The prefactor $i^{n^2}$ in \eqref{eq:Kaehlerpotential} is required, as follows from the Hodge-Riemann bilinear identities, to render the expression positive. We discuss these identities and their relevance for the positivity of the Weil-Petersson metric further in section \ref{app:positivity}.

Calabi-Yau manifolds are conjectured to exhibit mirror symmetry. This implies in particular the existence of a mirror manifold $\check X$ belonging to a deformation family $\check \cX \rightarrow \check \cM_{\text{ck}}$ varying over a space of complexified K\"ahler structures that can be identified with $\cX \rightarrow \cM_{\text{cs}}$. In the following, we will restrict to the case $n=3$. By recourse e.g.\ to the SYZ conjecture \cite{Strominger:1996it}, which for $n=3$ states that mirror symmetry can be understood as $T$-duality on the three directions of a 3-torus 
fibration, we see that a type IIA compactification on the manifold $X$ is identified with a type IIB compactification on the mirror manifold $\check X$. In type IIB compactifications, the vector multiplet moduli space is identified with the complex 
structure moduli space $\cM_{\text{cs}}$ of the compactification manifold, and the special K\"ahler base of the  hypermultiplet moduli space (over which the directions descending from the reduction of RR-forms are fibered) with the
complexified K\"ahler structure moduli space $\cM_{\text{ck}}$. For type IIA compactifications, this assignment is exchanged. 
Since for a large class of examples, the mirror pairs $(\cX,\check\cX)$ can be systematically constructed, see e.g.\  \cite{Batyrev:1993oya}, and the complex structure deformation spaces and their special K\"ahler structure are mathematically well understood, the discussion of the vector moduli spaces in this paper uses the 
identification with $\cM_{\text{cs}}$ (or $\check \cM_{\text{cs}}$), on whose structure we elaborate next.

\subsection[The complex structure moduli space of Calabi-Yau manifolds and the holomorphic $(3,0)$-form $\Omega$]{The complex structure moduli space of Calabi-Yau manifolds and the holomorphic \boldmath{$(3,0)$}-form \boldmath{$\Omega$}}
By the Calabi-Yau property, $H^{3,0}(\cX_z)$ is one-dimensional for all $z$. These one-dimensional vector spaces piece together over $\cM_{\text{cs}}$ to yield a complex line bundle $\cF^3$. Much of the study of $\cM_{\text{cs}}$ can be reduced to studying a holomorphic nowhere vanishing section $z \mapsto \Omega_z$ of this line bundle. The complexified middle cohomology of any fiber $\cX_z$ has a Hodge filtration
\begin{equation}
    H^3(\cX_z) = F^0 \supseteq F^1 \supseteq F^2 \supseteq F^3 = H^{3,0}(\cX_z) \,,
\end{equation}
where
\begin{equation}
    F^p = \bigoplus_{l \ge p} H^{l,3-l}(\cX_z) \,.
\end{equation}
Like $F^3$, $F^i$ for $i <3$ lift to holomorphic bundles $\cF^i$ over $\mathcal{M}_{\text{cs}}$. Local considerations of type imply that 
\begin{equation}
   \partial_{z_i} \Gamma(\mathcal{F}^k) \subseteq \Gamma(\mathcal{F}^{k-1}) \, .
\end{equation}
As $F^3$ is one-dimensional, one finds that
\begin{equation}
    \partial_{z_i} \Omega_z + (\partial_{z_i} K(z,\overline{z}))\Omega_z \in H^{2,1}(\cX_z) \,,
\end{equation}
motivating the definition of the covariant derivative
\begin{equation} \label{eq:covariant_derivative}
    \nabla_{z_i} = \partial_{z_i} + \partial_{z_i} K \, .
\end{equation}

A good strategy for studying $\Omega$ is to consider its periods, i.e.\ the integrals of $\Omega$ over cycles that are locally constant on moduli space. To this end, we note that $H_3(\cX_z,\mathbb{Z})$ is a symplectic module with respect to the intersection pairing. We can therefore introduce a symplectic basis of 3-cycles $\{A_I,B^I \,|\, I= 0,...,h^{2,1}\}$ that is constant over some contractible subset $U \subset \mathcal{M}_{\text{cs}}$. We then define the period vector
\begin{equation} \label{eq:period_vector}
    \Pi^T = \left(\int_{A_I} \Omega , \int_{B^I} \Omega\right) = (F_I, X^I)  \,.
\end{equation}
To interpret the entries of $\Pi$ as expansion coefficients of $\Omega$, we can introduce the fundamental classes $\{\eta_{A_I}, \eta_{B^I}\}$ of the cycles $\{A_I,B^I\}$, defined via the relations
\begin{equation}
    \int_{A_I} \gamma = \int \gamma \wedge \eta_{A_I} \,, \quad \int_{B_I} \gamma = \int \gamma \wedge \eta_{B_I} 
\end{equation}
for any closed 3-form $\gamma$. These classes satisfy the relation
\begin{equation}
    \int \eta_{A_I} \wedge \eta_{B^J} = A_I \cdot B^J = \delta_I{}^J \,.
\end{equation}
Hence,
\begin{equation} \label{eq:periods_of_etas}
    \int_{A_I} \eta_{B^J} = -\delta_{I}{}^J \,, \quad \int_{B^I} \eta_{A_J} = \delta^{I}{}_J \,,
\end{equation}
and all other periods of the fundamental classes vanish. From \eqref{eq:periods_of_etas}, we can deduce the relation between the fundamental classes and a dual basis  $\{\alpha^I, \beta_I \,|\, I= 0,...,h^{2,1} \}$ to the cycles $\{A_I,B^I\}$, defined via
\begin{equation} \label{eq:dual_basis}
    \int_{A_I} \alpha^J = \delta_{I}{}^J \,, \quad \int_{B^I} \beta_J = \delta^{I}{}_J \,,
\end{equation}
all other periods zero. It is
\begin{equation}
    \eta_{A_I} = \beta^I \,, \quad \eta_{B^I} = - \alpha_I \,.
\end{equation}
Finally, we can express $\Omega$ in either one of these sets of forms as
\begin{equation} \label{eq:Omega_expansion}
    \Omega = X^I \eta_{A_I} - F_I \eta_{B^I} = F_I \alpha^I + X^I \beta_I\,.
\end{equation}

\subsection{Positivity of the Weil-Petersson metric and of the gauge kinetic term} \label{app:positivity}
Identifying the local expansion coefficients $X^I$ and $F_I$ of the holomorphic (3,0)-form $\Omega$ as defined in \eqref{eq:Omega_expansion} with the entries of the local functions $\bX$, $\bF$ on target space introduced at the beginning of appendix \ref{app:4d_SUGRA}, the required positivity constraints on the Weil-Petersson metric and the gauge kinetic term follow from the Hodge-Riemann bilinear relations (see e.g.\ \cite{MR1288523}). These assert that for $\zeta$ a primitive  $(p,q)$ class in the middle cohomology of a complex $d$ dimensional K\"ahler manifold $M$, 
\begin{equation}
i^{(p-q)} (-1)^{d(d-1)/2} \int_M \zeta\wedge \bar \zeta > 0 \,.
\label{eq:Rblr}
\end{equation}
It is a classical result in the study of Riemann surfaces that these relations imply the positivity ${\rm Im}(\tau)>0$
for $\tau$ the holomorphic $g\times g$ matrix associated to the period matrix of a genus $g$ curve $\Sigma_g$: one introduces a basis $\{\omega_i \, | \, i=1,\ldots,g\}$ of $H^{1,0}(\Sigma_g)$ and a symplectic basis $\{a_i, b_i \,|\, i = 1 ,\ldots, g\}$ of $H_1(\Sigma_g,\IZ)$. The relations \eqref{eq:Rblr} then imply
\begin{equation}
    i \int \omega_i \wedge \bar{\omega_i} > 0 \,, \quad i,j=1,\ldots,g \,. \label{eq:hodge_riemann_10}
\end{equation}
Introducing the periods
\begin{equation}
    \tilde \Xi_{ij}=\int_{a_i}\omega_j \,, \quad \tilde \Phi_{ij}=\int_{b_i} \omega_j \,, \quad i,j=1,\ldots,g \,,
\end{equation} 
one defines $\tau$ (denoted $Z$ in \cite{MR1288523}) as the normalized $b_i$ period matrix,
\begin{equation} \label{eq:def_tau}
    \tau=\tilde \Phi \tilde \Xi^{-1} \,.
\end{equation}
A short calculation then shows that \eqref{eq:hodge_riemann_10} implies 
\begin{equation}
    \imP(\tau) > 0 \,.
\end{equation}
$\imP \tau$ plays the role of both the sigma model metric and the gauge coupling matrix in the context of rigid ${\cal N}=2$ supersymmetric theories, which were geometrized by Seiberg and Witten in their seminal paper \cite{Seiberg:1994rs}. 

The situation in $\cN=2$ supergravity theories in 4d, geometrized via compactifications of 10d theories on Calabi-Yau manifolds, is closely analogous. By the absence of cohomology in degree 1 on Calabi-Yau threefolds, \eqref{eq:Rblr} implies that
\begin{equation} \label{eq:hodge_riemann_30_12}
    i \int \Omega \wedge \bar{\Omega} > 0 \,, \quad - i \int \chi \wedge \bar{\chi} > 0 \quad \text{for} \quad \chi \in H^{2,1}(\cX_z) \,.
\end{equation}
Comparing to \eqref{eq:hodge_riemann_10}, we see that the role of the basis $\{\omega_i \,|\, i=1, \ldots, g \}$ of $H_1(\Sigma_g)$ is here played by $\{\Omega, \bar{\chi}_i \, | \, i = 1, \ldots, h^{2,1}\}$. The analogy between $\cN$ and $\tau$ is then immediately clear upon comparing \eqref{eq:sol_N} and \eqref{eq:def_tau}.\footnote{In terms of a prepotential $F$, the entries of the gauge coupling matrix $\tau$ in gauge theory are given by $\tau_{ij} = \partial_i \partial_j F$. The analogous quantity also exists in supergravity theories, and was introduced as $\boldsymbol{\cF}$ below \eqref{eq:NIJ}. In the context of Calabi-Yau compactifications, 
$\boldsymbol{\cN}$ and $\boldsymbol{\cF}$ are related to the Weil and Griffiths intermediate Jacobian respectively~\cite{MR0309937}, see \cite{MR2510071} for a review.}

Note that standard conventions in the physics literature (see e.g.\ \cite{MR2920151}) require $\imP(\tau) > 0$, but $\imP(\cN) < 0$. We have implemented the required sign by defining the $F_I$ as $A_I$ periods of $\Omega$ in \eqref{eq:period_vector}. A more common choice is to define $F_I$ as the negative $B^I$ periods.

In contradistinction to gauge theories, the sigma model metric and gauge coupling matrix do not coincide in supergravity theories. The former, given in terms of the K\"ahler potential in \eqref{eq:sigma_model_metric}, evaluates in the context of Calabi-Yau compactifications to
\begin{equation}
    g_{i\bar{\jmath}} = - \frac{\int \nabla_i \Omega \wedge \overline{\nabla_j \Omega}}{\int \Omega \wedge \bar{\Omega}} \,.
\end{equation}
The positivity of the metric follows immediately from \eqref{eq:hodge_riemann_30_12}.

To the best of our knowledge, the only known models satisfying the positivity constraint \eqref{eq:NIJ>0} are associated to Calabi-Yau geometries, or, at worst, to motives related to Calabi-Yau geometries (e.g.\ in the context of the mirrors of rigid Calabi-Yau manifolds \cite{Sethi:1994ch, Schwarz:1995ak} or of Calabi-Yau operators not necessarily associated to geometries \cite{MR3822913}). A proof that these constitute the only path towards constructing consistent $\cN=2$ supergravity models is currently out of reach;\footnote{A first modest step would be to extend the harmonicity argument in \cite{Seiberg:1994rs}, which demonstrates that $\imP \tau$ cannot be a global function on moduli space, to the supergravity setting.}
it would have far-reaching consequences for the $\cN=2$ version of the swampland program.

\subsection[Computing periods of $\Omega$ around the MUM point]{Computing periods of \boldmath{$\Omega$} around the MUM point} \label{app:computing_omega}
The most straightforward path towards computing the periods of $\Omega$ is as solutions to so-called Picard-Fuchs equations. For simplicity, we will restrict the following discussion to the case of $h^{2,1}=1$ (so-called one-parameter models). Since $\cF^0$ is 4-dimensional in this case, we find that $\Omega$ and its first four derivatives must be linearly dependent over the ring of holomorphic functions. This implies that the periods of $\Omega$ satisfy a fourth order differential equation 
\begin{equation} \label{eq:generic_PF}
    \left(f_0(z) + f_1(z) \theta + \ldots f_4(z) \theta^4 \right) \Pi = 0 \,, \quad \theta = z \frac{\dd}{\dd z} \, , 
\end{equation}
a so-called Picard-Fuchs equation (recall that the cycles $\{A_I, B^I\}$ were chosen to be locally constant on moduli space). This differential equation has special properties due to its geometrical origin, e.g.\ it is a Fuchsian differential equation with regular singularities lying in a compactification $\overline{\mathcal{M}_{\text{cs}}}$ of $\mathcal{M}_{\text{cs}}$. Due to these singular points, $\mathcal{M}_{\text{cs}}$ is not contractible; analytically continuing the periods $\Pi$ around singularities gives rise to multi-valued functions on $\mathcal{M}_{\text{cs}}$. The associated monodromy matrices reflect the ambiguity in a global choice of 3-cycles $\{A_I, B^I\} $ on all of $\mathcal{M}_{\text{cs}}$. 

The Frobenius method permits determining a basis of local solutions to \eqref{eq:generic_PF} around any given point $z_* \in \overline{\mathcal{M}_{\text{cs}}}$.
This Frobenius basis corresponds to periods with respect to some complex linear combinations of 3-cycles. A basis of solutions corresponding to an integral symplectic basis of 3-cycles is typically identified via a careful analysis of the geometry \cite{Candelas:1990rm}. However, mirror symmetry permits a simple algorithm for identifying such a basis, requiring only the knowledge of a few topological invariants of the mirror manifold $\check X$, provided there exists a point of maximal unipotent monodromy, a so-called MUM point, among the singular points in $\overline{\mathcal{M}_{\text{cs}}}$. At such a point, all indicial roots of \eqref{eq:generic_PF} are zero. If $z$ is chosen such that the MUM point is at $z=0$, the Frobenius basis takes the simple form
\begin{equation} \label{eq:Frobenius}
    \varpi =
    \begin{pmatrix*}[r]
        g_0(z) \\
        g_0(z)\log(z) + g_1(z) \\
        \frac{1}{2} g_0(z) \log^2(z) +g_1(z) \log(z) + g_2(z) \\
        \frac{1}{6} g_0(z) \log^3(z) + \frac{1}{2}g_1(z) \log^2(z) + g_2(z) \log(z) + g_3(z) 
    \end{pmatrix*}
\end{equation}
for power series $g_i(z)$ normalized by $g_0(0) = 1$ and $g_1(0)=g_2(0)=g_3(0)=0$. The monodromy of this period vector around the MUM point is captured by the matrix
\begin{equation}
    M^{\varpi}_0 = 
    \begin{pmatrix}
     1 & 0 & 0 & 0 \\
    2\pi  i & 1 & 0 & 0 \\
    \frac{(2\pi  i)^2}{2} & 2\pi  i & 1 & 0 \\
    \frac{(2\pi  i)^3}{6} & \frac{(2\pi  i)^2}{2} & 2\pi  i & 1 \\
    \end{pmatrix} \,.
\end{equation}
Mirror symmetry now dictates that the period vector around a MUM point in an appropriately chosen integral symplectic basis should have the general form
\begin{equation} \label{eq:period_at_MUM}
    \Pi = 
    \begin{pmatrix}
     F_0 \\
     F_1 \\
     X^0 \\
     X^1
    \end{pmatrix} =
    X^0 \begin{pmatrix}
    2 \cF(t) - t\partial_t \cF(t) \\
    \partial_t \cF(t) \\
    1 \\
    t
    \end{pmatrix} \, ,
\end{equation}
where $t(z) = \frac{1}{2\pi i} \log(z)+\frac{1}{2\pi i} \frac{g_1(z)}{g_0(z)}$ and 
\begin{equation}
    \cF(t) =-\frac{D\cdot D \cdot D}{6} t^3 - \frac{\sigma}{2}t^2  + \frac{c_2 \cdot D}{24}  t + \frac{\zeta(3) \chi }{2 (2 \pi i)^3}+  {\cF}_{\text{inst}}(e^{2 \pi i t})  \, .
    \label{eq:PrepotentialExpansion}
\end{equation}
Here $\chi$ is the Euler number of $\check X$, $D$ is a positive generator of $H_4(\check X, \mathbb{Z})$, $c_2$ is the second Chern class of $T\check X$ and $\sigma$ can be chosen to be 0 or $1/2$ depending on whether $D \cdot D \cdot D$ is even or odd. We have introduced the inhomogeneous prepotential $\mathcal{F}$, related to the homogeneous prepotential
\begin{equation} \label{eq:prepotential_from_periods}
    F(X) = \frac{1}{2} X^I F_I \,,
\end{equation}
with regard to which
\begin{equation}
    F_I = \partial_{X^I} F(X) \,,
\end{equation}
by 
\begin{align}
    \mathcal{F}(t) = \frac{1}{(X^0)^2}F(X) \,.
\end{align}

Due to the logarithmic structure of the periods at the MUM point, the mirror symmetry prediction \eqref{eq:PrepotentialExpansion} determines $\Pi$ uniquely up to a multiplicative constant. For suitably normalized $\Omega$, one thus finds
\begin{equation}
\label{eq:ChangeOfBasisMatrixAroundMUMPT}
    \Pi = (2\pi i)^3
    \begin{pmatrix}
    \frac{\zeta(3)\chi}{(2\pi i)^3} & \frac{c_2\cdot D}{24 (2\pi i)} & 0 & \frac{D \cdot D \cdot D}{(2\pi i)^3} \\
 \frac{c_2 \cdot D}{24} & -\frac{ \sigma}{2 \pi i} & -\frac{D\cdot D \cdot D}{(2\pi i)^2} & 0 \\
 1 & 0 & 0 & 0 \\
 0 & \frac{1}{2 \pi i} & 0 & 0
    \end{pmatrix} \varpi \, .
\end{equation}

In the case of multi-parameter models, the inhomogeneous prepotential (\ref{eq:PrepotentialExpansion}) generalizes to
\begin{equation} \label{eq:large_radius_F_multi_parameter}
    \cF =-\frac{\kappa_{ijk}}{6} t^i t^j t^k -\frac{\sigma_{ij}}{2}t^i t^j+  \gamma_j t^j + \frac{\zeta(3) \chi }{2 (2 \pi i)^3}+ {\cF}_{\text{inst}}(e^{2 \pi i t}) 
\end{equation}
and is likewise determined by topological data
 \begin{align} \label{eq:triple_intersection}
     \kappa_{ijk}&=D_i\cdot D_j\cdot D_k=\int_{\check X}\omega_i\wedge \omega_j \wedge \omega_k\,,\\ \gamma_k&=\frac{1}{24}c_2\cdot D_k=\frac{1}{24}\int_{\check X} c_2\wedge \omega_k\,, \\
     \chi&=\int_{\check X} c_3 \,.
 \end{align}
 Here, we have introduced a basis of $(1,1)$-forms $\omega_k$ dual to $D_k$. Note that this topological data, necessary to fix the perturbative contributions to \eqref{eq:large_radius_F_multi_parameter}, is precisely what is needed to fix the topological type of the Calabi-Yau threefold $\check X$ 
  by an application of the theorem of C.T.C. Wall~\cite{MR215313}. The coefficients $\sigma_{ij}$ have a less distinguished mathematical meaning. As we will now argue, their value can be inferred, up to choice of an integral symplectic basis, by the requirement that the monodromy group be contained in the integral symplectic group ${\rm Sp}(b_3,\mathbb{Z})$.

  Note that quadratic and lower order terms in the variables $t_i$ 
  do no affect the Yukawa couplings, and do not affect the Weil-Peterssen metric if they are real. The quadratic terms do however enter, as we discuss in section \ref{subsubsec:T_in_4d_inst}, in the topological gauge coupling $\reP{\cal N}_{ij}$ and therefore play a role when discussing time reversal invariance.  
 
 Let us consider the shift monodromy $T_i:t^i\mapsto t^i+1$ defined by analytic continuation around the divisor 
 $\{z_i=0\}\subset {\cal M}_{\text{cs}}$, where
 \begin{equation}
 t^i=\frac{1}{2 \pi i}\log(z_i)+O(z)\,, \quad i=1,\ldots,r\,, \quad r=h^{2,1}(X)=h^{1,1}(\hat X) \,.
 \end{equation}
 It acts on the period vector
 \begin{equation} 
    \Pi = 
    \begin{pmatrix}
     F_0 \\
     F_1 \\
     \vdots\\
     F_r \\
     X^0 \\
     X^1 \\
     \vdots \\
     X^r
    \end{pmatrix} =
    X^0 \begin{pmatrix}
    2 \cF(t) - t^i\partial_{t^i} \cF(t) \\
    \partial_{t^1} \cF(t) \\
    \vdots \\
    \partial_{t^r} \cF(t) \\
    1 \\
    t^1 \\
    \vdots \\
    t^r
    \end{pmatrix} =
        X^0 \begin{pmatrix}
     \frac{\kappa_{ijk}}{6} t^i t^j t^k +  \gamma_j t^j + \frac{\zeta(3) \chi }{(2 \pi i)^3}+ \ldots  \\
    -\frac{\kappa_{1jk}}{2} t^j t^k -\sigma_{1j} t^j+  \gamma_1 +\ldots \\
    \vdots \\
        -\frac{\kappa_{rjk}}{2} t^j t^k -\sigma_{rj} t^j+  \gamma_r +\ldots  \\
    1 \\
    t^1 \\
    \vdots \\
    t^r
    \end{pmatrix} 
\end{equation}
as\footnote{In these $(2r+2)\times (2r +2)$-matrices 
 $T_i$, the $\underline{0}$ and $\underline{0}^T$ represent rows and columns of $r$ zeros. $\delta_{i \boldsymbol{\cdot}}$ is a row vector with entries $\delta_{ij}$, $\delta_{\boldsymbol{\cdot}i}$ is its transpose. Likewise, $\kappa_{ii \boldsymbol{\cdot}}$ and $\sigma_{i \boldsymbol{\cdot}}$ are row vectors with entries $\kappa_{iij}$ and $\sigma_{ij}$ respectively, with respective transposes $\kappa_{\boldsymbol{\cdot}ii} $ and $\sigma_{\boldsymbol{\cdot}i}$. The boldface symbols represent $r\times r$ blocks of entries.}   (here we use the same notation for the shift monodromy and its representation on the space of periods)   
 \begin{equation}
T_i=\left(\begin{array}{cccc}
1&\ \ \ -\delta_{i\boldsymbol{\cdot}}\ \ & \frac{1}{6}\kappa_{iii}+ 2 \gamma_i&\ \  \frac{1}{2}\kappa_{ii\boldsymbol{\cdot}}+ \sigma_{i\boldsymbol{\cdot}}\\
\underline{0}^T& \boldsymbol{1}& \sigma_{\boldsymbol{\cdot}i}-\frac{1}{2} \kappa_{\boldsymbol{\cdot}ii}& -\boldsymbol{\kappa}_{i\boldsymbol{\cdot}\boldsymbol{\cdot}}\\ 
0& \underline{0}& 1 & \underline{0} \\
\underline{0}^T& \bf{0}& \delta_{\boldsymbol{\cdot}i} &\boldsymbol{1}\\ 
\end{array}\right)\, .     
\label{eq:TshiftsatMUM}
 \end{equation}

The fractions that appear in this matrix render its integrality non-trivial. The integrality of the entry $(T_i)_{1,r+2}$ is guaranteed by the fact that it computes the arithmetic genus of the divisor $D_i$, see e.g.~\cite{Witten:1996bn}: as $c_1(\hat X)=0$,
\begin{equation} 
\mathbb{Z}\ni\chi(D_i,{\cal O}_{D_i})=\int_{\hat X} (1-e^{-D}) {\rm td}(\hat X)= \frac{1}{12}(2 D_i^3+ c_2\cdot D_i)=\frac{1}{6}\kappa_{iii} + 2 \gamma_i=(T_i)_{1,r+2}\, .
\end{equation}
The remaining obstructions to integrality are the elements $\frac{1}{2}\kappa_{ii\boldsymbol{\cdot}}+ \sigma_{i\boldsymbol{\cdot}}$ and $\sigma_{\boldsymbol{\cdot}i}-\frac{1}{2} \kappa_{\boldsymbol{\cdot}ii}$. As $\kappa_{ijk}$ is symmetric in all of its indices, integrality of these elements follows from the choice
\begin{equation} \label{eq:cond_sigma}
    \sigma_{ij} = \frac{\kappa_{iij}}{2} + n_{ij} \,, \quad n_{ij} \in \IZ \,.
\end{equation}
One can easily check that all symmetric choices \eqref{eq:cond_sigma} (i.e.\ $n_{ij} = n_{ji}$) lead to symplectic matrices $T_i$, e.g.\ by first checking this condition at vanishing $n_{ij}$, and then shifting the $\sigma_{ij}$ by a choice of integers $n_{ij}$ via multiplication of $T_i$ with the matrix 
\begin{equation}
\left(\begin{array}{cccc}
1&\ \ \ \underline{0} \ \ & -n_{ii} &\ \  n_{i\boldsymbol{\cdot}}  \\
\underline{0}^T& \boldsymbol{1}& n_{\boldsymbol{\cdot}i}  & \boldsymbol{0}\\ 
0& \underline{0}& 1 & \underline{0} \\
\underline{0}^T& \bf{0}& \underline{0}^T  &\boldsymbol{1}\\ 
\end{array}\right)   
 \end{equation}
which is clearly integer symplectic if $n$ is symmetric. 

Note that
the $K$-theory class of a geometric D$4$ brane calculates the $\sigma_{ij}$ geometrically, see e.g.\ 
\cite{Huang:2006hq}\cite{Gerhardus:2016iot}. But also in this setting, a basis of all branes has to be chosen
such that the auto equivalences in the derived category of $B$-branes are integer symplectic, resulting in the same integral freedom in choosing the $\sigma_{ij}$.
   
Finally, to check the consistency of our sign conventions in \eqref{eq:Omega_expansion} and \eqref{eq:large_radius_F_multi_parameter}, we will compute the sign of $\int_X \Omega \wedge \bar{\Omega}$ in the limit of large $v^i = \imP t^i$, where
\begin{equation}
    \cF \sim -\frac{\kappa_{ijk}}{6} t^i t^j t^k \,.
\end{equation}
By \eqref{eq:Omega_expansion},
\begin{align}
   i \int_X \Omega \wedge \bar{\Omega} &=i( F_I \bar{X}^I - X^I \bar{F}_I) \\
    &= -2 |X^0|^2 \imP \left( \frac{\kappa_{ijk}}{6} t_i t_j t_k - \frac{\kappa_{ijk}}{2} \bar{t}_i t_j t_k \right) \nn\\
    &= \frac{4}{3} |X^0|^2 \kappa_{ijk} v^i v^j v^k \,. \nn
\end{align}
For $v^i > 0$, this is manifestly positive. For the canonical orientation of $X$ in which the volume form is locally proportional to $( i \dd z^1 \wedge \dd \bar{z}^1) \wedge (i \dd z^2 \wedge \dd \bar{z}^2) \wedge (i \dd z^3 \wedge \dd \bar{z}^3)$, this is as it should be, in accord with the Hodge-Riemann bilinear relations \eqref{eq:Rblr}.

\section{Supersymmetric flux vacua and rank 2 attractors}
\label{appendix:SUSY_vacua}
In this appendix, we consider Calabi-Yau compactifications of type IIB supergravity in the presence of non-trivial fluxes $F_3$ and $H_3$. We will consider backgrounds with the axio-dilaton
\begin{equation}
    \tau = C_0 + i e^{-\phi} \,,
\end{equation}
constant over the internal dimensions, with $\imP \tau = e^{-\phi} > 0$. The quantities which determine the locus of vacua in this class of theories are the K\"ahler potential ${\cal K}$ of the 
full physical theory 
\begin{equation}  
{\cal K}=-\log(i\int_X\Omega\wedge\bar \Omega)- \log(-i(\tau - \bar \tau)) + K_{\text{hyper}}
\label{eq:fullKaehler}
\end{equation}
and the superpotential
\begin{equation}
    W = \int_X G_3 \wedge \Omega \,,
\end{equation}
which we have written in terms of the convenient quantity
\begin{equation}
    G_3 = F_3 - \tau H_3 \,.
\end{equation}
Together, these determine the 4d potential
\begin{equation} \label{eq:sugra_V}
    V = e^{\cK} \left( \sum g^{\indHVI \bar{\indHVII}} \nabla_\indHVI W \overline{\nabla_\indHVII W} - 3 |W|^2 \right) \,,
\end{equation}
at the minima of which the vacua lie. The indices $\indHVI, \indHVII$ here run over all complex scalar fields descending from the vector and hypermultiplets,\footnote{Slightly abusing terminology, as we are now considering $\cN=1$ supersymmetry, we will continue to call these vector multiplet scalars and hyperscalars, respectively.} including the axio-dilaton $\tau$ contained in the universal hypermultiplet. The symbol $\nabla_{\Xi}$ indicates the covariant derivative ${\nabla_{\Xi} = \partial_{\Xi} + \partial_{\Xi}\cK}$.

At tree level in $\alpha'$ and $g_s$, the only dependence of the superpotential on the hypermultiplet sector is via $\tau$.  Hence, 
\begin{equation} \label{eq:leading_hypers_to_V}
    \sum g^{i \bar{\jmath}} \nabla_i W \overline{\nabla_j W}  = \sum g^{i \bar{\jmath}} \cK_i \cK_{\bar{\jmath}} |W|^2 \,,
\end{equation}
with the sum extending over all geometric hyperscalars save the axio-dilaton.\footnote{Neither $W$ nor $\cK$ depend on the hyperscalars descending from the RR-sector. These hence do not contribute to the potential $V$.}  Again at leading order, the contribution \eqref{eq:leading_hypers_to_V} cancels the second term in parentheses in \eqref{eq:sugra_V} due to the form of $K_{\text{hyper}}$ \cite{Giddings:2001yu}, giving rise to a so-called no-scale potential, as $V$ is independent of the hyperscalars which parametrize the K\"ahler structure of the compactification Calabi-Yau manifold. The remaining terms are of the form
\begin{equation}
    V = e^{\cK} \sum_{A,B} g^{A \bar{B}} \nabla_A W \overline{\nabla_B W} \,,
\end{equation}
with the sum now extending over all vector multiplet scalars as well as the axio-dilaton $\tau$. Vacua lie at the minimum of the potential, $V=0$, hence require 
\begin{equation} \label{eq:supocrit}
    \nabla_A W = 0  \,.
\end{equation}
As the hyperscalars are uncharged, the VEV of the supersymmetry variations of the associated fermions are proportional to $\nabla_i W$. A vacuum preserving supersymmetry must thus satisfy, in addition to \eqref{eq:supocrit}, 
\begin{equation}
    0 = \nabla_i W = \cK_i W \quad \Rightarrow \quad W=0 \,.
\end{equation}

Since $\langle [\nabla_{a}\Omega] \,| \,a \in \text{vector multiplet index set} \rangle = H^{2,1}(\cX_{z_0})$, 
(\ref{eq:supocrit}) implies that the projection of $G_3$ into $H^{1,2}(\cX_{z_0})$ must vanish. Recalling that we are considering backgrounds with $\tau$ constant on the internal dimensions, the computation
\begin{equation} \label{eq:dtauW}
    \nabla_\tau W = - \int H_3 \wedge \Omega + W \partial_\tau \cK = - \int H_3 \wedge \Omega - \frac{\int G_3 \wedge \Omega}{\tau - \bar{\tau}} = \frac{-1}{\tau - \bar{\tau}} \int \overline{G_3} \wedge \Omega \,,
\end{equation}
shows that $ \nabla_{\tau} W=0$ implies the vanishing of the projection of $G_3$ into $H^{3,0}(\cX_{z_0})$. $V=0$ hence holds for points $z_0$ in $\cM_{\text{cs}}$ such that
\begin{equation} \label{eq:vac_constraint_on_G3}
G_3 \in H^{2,1}(\cX_{z_0})\oplus H^{0,3}(\cX_{z_0})\ .
\end{equation}    
Imposing supersymmetry, the vanishing of \eqref{eq:dtauW} imposes the stronger condition
\begin{equation} \label{eq:H_wedge_Omega}
     \int H_3 \wedge \Omega = 0 \,.
\end{equation}
As $H_3$ is real, $z_0$ is hence constrained to satisfy
\begin{equation} \label{eq:susy_condition_H3}
    H_3 \in H^{2,1}(\cX_{z_0}) \oplus H^{1,2}(\cX_{z_0}) \,.
\end{equation}
The constraint $W=0$ together with \eqref{eq:H_wedge_Omega} implies
\begin{equation}
    \int F_3 \wedge \Omega = 0 \,,
\end{equation}
thus also
\begin{equation} \label{eq:susy_condition_F3}
    F_3 \in H^{2,1}(\cX_{z_0}) \oplus H^{1,2}(\cX_{z_0}) \,.
\end{equation}

A detailed study of the equations of motion of type IIB supergravity resulting from a warped compactification ansatz shows that they cannot be satisfied unless the total tension of all localized sources is negative \cite{Maldacena:2000mw, Giddings:2001yu}. Choosing these sources to be D$3$ branes, D$7$ branes and O$3$ planes, the BPS condition implies that the resulting total D$3$ brane charge $Q_3$ is also negative. We will now argue that for Calabi-Yau compactifications, this does not give rise to an additional constraint on the fluxes $F_3$ and $H_3$. The Bianchi identity for the 5-form field strength $\tilde{F}_5$ in the presence of localized sources is given by\footnote{Note that the relative sign between the two terms in equation \eqref{eq:F_5_bianchi} is different from the one in reference \cite{Giddings:2001yu}. We follow the conventions of \cite{Blumenhagen:2013fgp}: the relative sign in \eqref{eq:F_5_bianchi} follows from the equations of motion for $F_5$ upon imposing self-duality, see equation (16.179) in \cite{Blumenhagen:2013fgp}, or the derivation of $N_{\text{flux}}$ leading up to (17.50).}
\begin{equation} \label{eq:F_5_bianchi}
d\tilde F_5=H_3\wedge F_3-2 \kappa_{10}^2 T_3 \delta_{\text{loc}} \,.
\end{equation}
Here, $T_3$ indicates the D$3$ brane tension, and $\delta_{\text{loc}}$ signifies the Poincaré dual to the worldvolume of the sources. Integrating, we obtain
\begin{equation}
\int_X F_3\wedge H_3 =- 2 \kappa_{10}^2 T_3 Q_3  \,,
\end{equation} 
with $Q_3$ denoting the localized D$3$ brane charge, carried e.g.\ by D$3$ branes, D$7$ branes and O$3$ planes. We therefore find that any solution satisfying the warped compactification ansatz requires \cite{Kachru:2020sio} 
\begin{equation} \label{eq:crossed_fluxes}
    \int F_3 \wedge H_3 > 0 \,.
\end{equation}
This condition does not pose an independent constraint on fluxes leading to supersymmetric solutions however, as it follows from \eqref{eq:susy_condition_H3}, \eqref{eq:susy_condition_F3}, and reality of the dilaton:
\begin{equation}
    \int F_3 \wedge H_3 = \frac{e^{\phi}}{2i} \int G_3 \wedge \bar{G}_3 > 0 \,,
\end{equation}
the inequality following from the Hodge-Riemann bilinear identities \eqref{eq:Rblr} for primitive (2,1)-forms.

Changing the focus from a choice of fluxes to a choice of distinguished points on moduli space, we arrive at the following statement: a point $z_0$ in the complex structure moduli space corresponds to a supersymmetric flux vacuum if and only if there exists a rank 2 lattice $\Gamma \subset H^3(\cX_{z_0},\mathbb{Z})$ such that the complexification $\Gamma_{\mathbb{C}} = \Gamma \otimes_{\mathbb{Z}} \mathbb{C}$ satisfies
\begin{equation} \label{eq:hodge_lambda}
    \Gamma_{\mathbb{C}} = (\Gamma_{\mathbb{C}} \cap H^{2,1}(\cX_{z_0})) \oplus (\Gamma_{\mathbb{C}} \cap H^{1,2}(\cX_{z_0})) \, .   
\end{equation}
To see this, note that given $\Gamma = \langle \gamma_1, \gamma_2 \rangle_{\IZ}$ satisfying \eqref{eq:hodge_lambda}, constants $\alpha_{1,2} \in \IC$ exist such that 
\begin{equation}
    \Gamma_{\mathbb{C}} \cap H^{2,1}(\cX_{z_0})) = \langle \alpha_1 \gamma_1 - \alpha_2 \gamma_2 \rangle_{\IC}\,.
\end{equation}
Without loss of generality, we will assume that $\imP \frac{\alpha_2}{\alpha_1} > 0$. Up to positive integral multiples, the following four options for the choice of $G_3 \in H^{2,1}(\cX_{z_0})$, with the corresponding choice of $F_3$, $\tau$, and $H_3$ indicated by the successive parentheses, are then consistent with real string coupling (i.e.\ $\imP \tau > 0$):
\begin{align} \label{eq:4_choices_to_identify_G3}
    &(\gamma_1) - (\frac{\alpha_2}{\alpha_1})(\gamma_2) &,&  &(\gamma_2) - (-\frac{\alpha_1}{\alpha_2})(-\gamma_1)\,, \nn \\
    &(-\gamma_1) - (\frac{\alpha_2}{\alpha_1})(-\gamma_2) &,&  &(-\gamma_2) - (-\frac{\alpha_1}{\alpha_2})(\gamma_1)\,.
\end{align}
To ensure the positivity of the string coupling, we have the freedom to distribute a factor of $N \in \IN$ in our identification of $\tau$ and $H_3$,
\begin{equation}
     \tau H_3 \rightarrow (\frac{\tau}{N})(N H_3) \,.
\end{equation} 

Specializing to the case of one-parameter models, i.e.\ $b^3=4$, we see that the lattice
\begin{equation}
    \Lambda = \left(H^{3,0}(\cX_{z_0}) \oplus H^{0,3}(\cX_{z_0})\right) \cap H^{3}(X,\IZ) 
\end{equation}
orthogonal to $\Gamma$ in $H^3(\cX_{z_0},\IZ)$ (with regard to the inner product induced by integration of the wedge product of forms) must in this case also have rank 2. A point $z_0$ for which $\Lambda$ has rank 2 is called a rank 2 attractor, due to its relation to black hole solutions in 4d $\cN=2$ supergravity theories \cite{Ferrara:1995ih}.
Over $\IQ$, the two lattices $\Lambda$ and $\Lambda^\perp = \Gamma$ generate the third cohomology, 
\begin{equation} 
H^3(\cX_{z_0},\mathbb{Q})=\Lambda_{\mathbb{Q}} \oplus \Lambda^\perp_{\mathbb{Q}} \,.
\end{equation}

\section{List of rank 2 attractor points and a strategy to find them}
\label{appendix:Finding rank 2 attractors}
Given a one-parameter family $\cX$ of Calabi-Yau threefolds we want to find rank two attractor points (or equivalently points admitting supersymmetric flux vacua), i.e.\ complex structure parameters $z_0$ such that
\begin{align}
    H^3(\cX_{z_0}(\mathbb{C}),\mathbb{Q}) = \Lambda \oplus \Lambda^\perp
    \label{eq:HodgeSplitting}
\end{align}
for 
\begin{align}
    \Lambda \subset H^{3,0}(\cX_{z_0}(\mathbb{C})) \oplus H^{0,3}(\cX_{z_0}(\mathbb{C})) \quad \text{and} \quad \Lambda^\perp \subset H^{2,1}(\cX_{z_0}(\mathbb{C})) \oplus H^{1,2}(\cX_{z_0}(\mathbb{C}))\, .
\end{align}
A beautiful method for doing this was given in \cite{Candelas:2019llw}. The idea is as follows. If $\cX_{z_0}$ is defined over some number field $K$, one can for suitable primes $p$ reduce to a variety $\cX_{z_0,p}=X_{z_0} / \mathbb{F}_p$ defined over the finite field with $p$ elements. The Frobenius automorphism $F_p : x \mapsto x^p$ of $\overline{\mathbb{F}_p}$ then naturally acts on $\cX_{z_0,p}(\overline{\mathbb{F}_p})$ with the fixed points being exactly the points over $\mathbb{F}_p$. For suitable $p$-adic cohomology groups $H^3(\cX_{z_0,p})$ the Frobenius automorphism induces an action $F_p^*$ on $H^3(\cX_{z_0,p})$ and Hodge-like conjectures suggest that a splitting of the form (\ref{eq:HodgeSplitting}) induces a splitting of $H^3(\cX_{z_0,p})$ which is compatible with the action of $F_p^*$. Practically speaking, this gives a factorization of the Frobenius polynomial
\begin{align}
    \det(1-T F_p^* |H^3(\cX_{z_0,p})) = (1-a_pT+p^3T^2)(1-b_ppT+p^3T^2)
    \label{eq:FrobeniusFactorization}
\end{align}
with integers $a_p$ and $b_p$. Conversely, Tate-like conjectures suggest that such a splitting for enough primes $p$ leads to the splitting (\ref{eq:HodgeSplitting}). This gives the following strategy for finding rank two attractor points:
\begin{itemize}
    \item[(1)] Compute the Frobenius polynomials for some primes $p$ and all smooth distinct fibers of $\cX$ (which are finitely many after the reduction to $\mathbb{F}_p$). 
    \item[(2)] Look for persistent factorizations of the Frobenius polynomial and try to reconstruct the underlying complex structure parameters lying in some number field $K$. 
\end{itemize}
For an efficient algorithm to compute the Frobenius polynomials directly from the Picard-Fuchs operator see e.g.\ \cite{candelas2021}. 

The bigger picture of the splitting (\ref{eq:FrobeniusFactorization}) is that 4-dimensional representations of $\text{Gal}(\overline{K}/K)$ split to sums of two 2-dimensional representations. E.g.\ for $K=\mathbb{Q}$ the latter representations are well understood and in our case it is known that they they are associated with Hecke eigenforms of weight 4 and 2. These can be identified from the knowledge of the Frobenius polynomials and conjecturally all periods of $\cX_{z_0,p}$ can be expressed in terms of the periods and quasiperiods of these Hecke eigenforms. This is worked out for some examples in \cite{BoenischThesis} and \cite{ADEK}. 

In Table \ref{tab:attractor} we list some rank two attractor points defined over $\mathbb{Q}$ with the associated Hecke eigenforms. This includes two hypergeometric models and ten models where the attractor point is an apparent singularity which is fixed under an involutional symmetry $z \mapsto \frac{1}{a^2z}$ which induces the splitting. For apparent singularities the deformation method from \cite{candelas2021} still works if one slightly modifies the Wronski matrix used in the computation such that it becomes invertible at the apparent singularity. 

\newcolumntype{R}{>{$}r<{$}}
\newcolumntype{L}{>{$}l<{$}}
\newcolumntype{M}{R@{${}={}$}L}

\begin{table}[h]
    \centering
    \footnotesize
    \begin{tabular}{c|c|c|c|M|M}
  Operator & $z_0$ & $N_f$ & $N_g$ & \multicolumn{2}{c|}{$a_p$} & \multicolumn{2}{c}{$b_p$}\\ \hline
4    & $-1/2^33^6$       & 54  & 54  & a_5                   & 3                     & a_5                  & 3         \\
11   & $-1/2^43^3$       & 180 & 36  & (a_7,a_{11})          & (2,30)                & \multicolumn{2}{c}{} \\
34   & $-1/7$            & 14  & 14  & a_3                   & 8                     & \multicolumn{2}{c}{} \\
36   & $-1/2^6$          & 96  & 32  & (a_5,a_7)             & (2,12)                & \multicolumn{2}{c}{} \\
49   & $-1/2^45$         & 400 & 400 & (a_3,a_{7})           & (4,-16)               & (a_3,a_{7})          & (-2,2)    \\
55   & $1/2^4$           & 60  & 20  & a_7                   & -28                   & \multicolumn{2}{c}{} \\
84   & $-1/2^4$          & 20  & 20  & \multicolumn{2}{c|}{} & \multicolumn{2}{c}{} \\
84   & $1/2^43$          & 12  & 36  & \multicolumn{2}{c|}{} & \multicolumn{2}{c}{} \\
100  & $1/2^3$           & 14  & 14  & a_3                   & -2                    & \multicolumn{2}{c}{} \\
101  & $1$               & 22  & 11  & a_3                   & -7                    & \multicolumn{2}{c}{} \\
103  & $-1/3^2$          & 180 & 90  & (a_7,a_{11})          & (-28,24)              & (a_7,a_{11})         & (-4,0)    \\
107  & $-1/2^5$          & 48  & 48  & a_5                   & 6                     & \multicolumn{2}{c}{} \\
111  & $-1/2^83$         & 144 & 144 & a_5                   & -14                   & a_5                  & 2         \\
115  & $-1/2^8$          & 32  & 32  & a_3                   & 8                     & \multicolumn{2}{c}{} \\
144  & $-1/2^33^2$       & 306 & 306 & a_5                   & 12                    & (a_5,a_7,a_{23})     & (0,2,6)   \\
145  & $-1/3^6$          & 108 & 54  & (a_5,a_7)             & (9,-1)                & a_5                  & -3        \\
154  & $-1/2^43^3$       & 216 & 216 & a_5                   & 4                     & a_5                  & 4         \\
155  & $-1/2^{12}$       & 128 & 128 & (a_3,a_5)             & (-2,-6)               & (a_3,a_5)            & (-2,2)    \\
165  & $-1/3^3$          & 54  & 27  & a_5                   & 12                    & \multicolumn{2}{c}{} \\
166  & $-1/2^83^6$       & 864 & 864 & (a_5,a_7,a_{11})      & (-19,13,65)           & (a_5,a_7,a_{11})     & (1,-3,-3) \\
238  & $-1/2^4$          & 88  & 88  & a_3                   & -1                    & a_3                  & -3        \\
277  & $-1/2^{14}$       & 240 & 80  & (a_7,a_{11})            & (28,24)               & a_3                  & 2         \\
2.32 & $1/2^43^3$        & 324 & 324 & a_5                   & -3                    & (a_5,a_7)            & (3,2)
\end{tabular}
    \caption{Some rational rank two attractor points of Calabi-Yau operators. The labels corresponds to the number in \cite{almkvist2005} and in the last case to the number from \cite{CYDatabase}. We give enough Hecke eigenvalues $a_p$ and $b_p$ to specify the associated normalized Hecke eigenforms $f \in S_4(\Gamma_0(N_f))^{\text{new}}$ and $g \in S_2(\Gamma_0(N_g))^{\text{new}}$ uniquely.}
    \label{tab:attractor}
\end{table}

We close this section by giving some more complicated rank two attractor points of operators from the AESZ list \cite{almkvist2005}. The first examples of attractor points not defined over $\mathbb{Q}$ are the points $33 \pm 8 \sqrt{17}$ of the operator AESZ 34. These were found in \cite{Candelas:2019llw} and are again associated with Hecke eigenforms of weight 4 and 2. Another example are the points $\pm \sqrt{-1}/2^4$ of AESZ 105. There the Galois representation on $\Lambda^\perp$ is that of the elliptic curve $E: \, y^2=x^3+(1\pm\sqrt{-1})x^2+1$ (after a Tate twist) or equivalently that of the weight 2 Bianchi modular form associated with $E$. Numerically one finds that the periods of $\Lambda^\perp$ and $E$ (multiplied by $2\pi i$) agree. We expect that $\Lambda$ is related to a weight 4 Bianchi modular form but we have not tried to identify this form. The same holds for the points $\pm \sqrt{-3} /3^2$ of AESZ 161 which are associated with the elliptic curve $y^2+xy+(a+1)y=x^3-x^2+(a-8)x+3a-8$ with $a=\frac{1 \pm \sqrt{-3}}{2}$.

\bibliographystyle{JHEP.bst}
\bibliography{References}

\end{document}